\documentclass{aa}
\usepackage{graphicx}
\usepackage{xcolor}
\usepackage{txfonts}
\begin{document}

\authorrunning{K. Burdonov et al.,}
\titlerunning{Inferring possible magnetic field strength of accreting inflows in EXor-type objects from scaled laboratory experiments}
   \title{Inferring possible magnetic field strength of accreting inflows in EXor-type objects from scaled laboratory experiments}

   \author{K. Burdonov\inst{1,2,3}, R. Bonito\inst{4}, T. Giannini \inst{5}, N. Aidakina\inst{3}, C. Argiroffi\inst{4,6}, J. Béard\inst{7}, S.N. Chen\inst{8}, A. Ciardi\inst{2},  V. Ginzburg\inst{3}, K. Gubskiy\inst{9}, V. Gundorin\inst{3}, M. Gushchin \inst{3},  A. Kochetkov\inst{3}, S. Korobkov\inst{3}, A. Kuzmin\inst{3}, A. Kuznetsov\inst{9}, S. Pikuz\inst{9,10}, G. Revet\inst{1}, S. Ryazantsev\inst{9,10}, A. Shaykin\inst{3}, I. Shaykin\inst{3}, A. Soloviev\inst{3}, M. Starodubtsev\inst{3}, A. Strikovskiy\inst{3}, W. Yao\inst{1,2}, I. Yakovlev\inst{3}, R. Zemskov\inst{3}, I. Zudin\inst{3}, E. Khazanov\inst{3}, S. Orlando\inst{4}, and J. Fuchs\inst{1}
          }

   \institute{LULI - CNRS, CEA, UPMC Univ Paris 06 : Sorbonne Université, Ecole Polytechnique, Institut Polytechnique de Paris - F-91128 Palaiseau cedex, France
   \and
   Sorbonne Université, Observatoire de Paris, PSL Research University, LERMA, CNRS UMR 8112, F-75005, Paris, France
   \and
   IAP, Russian Academy of Sciences, 603950, Nizhny Novgorod,  Russia
   \and
   INAF - Osservatorio Astronomico di Palermo, Piazza del Parlamento 1, 90134 Palermo, Italy
   \and
   INAF - Osservatorio Astronomico di Roma, Via Frascati 33, 00078 Monteporzio Catone, Italy
   \and   
   Department of Physics and Chemistry, University of Palermo, 90133, Palermo, Italy
   \and
   LNCMI, UPR 3228, CNRS-UGA-UPS-INSA, 31400, Toulouse, France
   \and
   ELI-NP, ”Horia Hulubei” National Institute for Physics and Nuclear Engineering, 30 Reactorului Street, RO-077125, Bucharest-Magurele, Romania
   \and
   National Research Nuclear University “MEPhI”, 115409, Moscow, Russia
   \and
   Joint Institute for High Temperatures, RAS, 125412, Moscow, Russia
   }
    
   \date{Accepted for publication in $Astronomy$ $\text{\&}$ $Astrophysics$ on February 8, 2021}

% \abstract{}{}{}{}{} 
% 5 {} token are mandatory
 
  \abstract
   {}
   {EXor-type objects are protostars that display powerful UV-optical outbursts caused by intermittent and powerful events of magnetospheric accretion. These objects are not yet well investigated and are quite difficult to characterize. Several parameters, such as plasma stream velocities, characteristic densities, and temperatures, can be retrieved from present observations. As of yet, however, there is no information about the magnetic field values and the exact underlying accretion scenario is also under discussion.}
   {We use laboratory plasmas, created by a high power laser impacting a solid target or by a plasma gun injector, and make these plasmas propagate perpendicularly to a strong external magnetic field. The propagating plasmas are found to be well scaled to the presently inferred parameters of EXor-type accretion event, thus allowing us to study the behaviour of such episodic accretion processes in scaled conditions.}
   {We propose a scenario of additional matter accretion in the equatorial plane, which claims to explain the increased accretion rates of the EXor objects, supported by the experimental demonstration of effective plasma propagation across the magnetic field. In particular, our laboratory investigation allows us to determine that the field strength in the accretion stream of EXor objects, in a position intermediate between the truncation radius and the stellar surface, should be of the order of 100 gauss. This, in turn, suggests a field strength of a few kilogausses on the stellar surface, which is similar to values inferred from observations of classical T Tauri stars.}
   {}


\maketitle
%
%________________________________________________________________

\section{Introduction}
\label{sec:intro}

Low-to-intermediate mass protostars (0.1-8 M$_\odot$) accrete their mass from the material inside the circumstellar disc. About 90\% of the final mass is accreted onto the star in about 10$^{5-6}$ yr, with typical mass accretion rates of 10$^{-7}-10^{-5}$ M$_\odot$/yr (main accretion phase). In the subsequent 10$^7$ yr, the accretion progressively fades to rates of 10$^{-10}-10^{-9}$ M$_\odot$/yr (classical T Tauri stars; CTTSs), until the star reaches the main-sequence evolutionary track.

Although small and irregular photometric variations ($\Delta$V$\sim$0.1$-$1 mag) caused by disc accretion variability are commonly observed in CTTSs, a few dozen of the young sources display powerful UV-optical outbursts of much larger intensity (up to $4-7$ mag). These outbursts are caused by intermittent and powerful events of magnetospheric accretion (\citealt{1994ApJ...429..797S}). Historically, these objects were serendipitously found during observational campaigns dedicated to different scientific aims. From an observational point of view, protostellar eruptive variables are classified in two main groups (\citealt{2014prpl.conf..387A}). The first group are FU Orionis objects (or FUors, a class defined after the prototype source FU Ori),  which are characterized by bursts with $\Delta$V$\sim$6$-$8 mag, duration of decades, accretion rates of 10$^{-5}-10^{-4}$ M$_\odot$/yr, and spectra dominated by absorption lines. The second group are EX Lupi objects (or EXors, a class defined after the prototype source EX Lup), which are characterized by less powerful outbursts ($\Delta$V$\sim$3$-$5 mag) with duration of months to one year, recurrence time of months to years, accretion rates of 10$^{-7}-10^{-6}$ M$_\odot$/yr, and emission line spectra. In addition, in the last decade a handful of objects have been found that show outbursts with amplitude and timescales in between those of classical EXors and FUors (e.g. HBC 722 and V1647 Ori, \citealt{2014prpl.conf..387A} and references therein).
In the last decade several multiwavelength sky surveys
(e.g. Gaia\footnote{https://sci.esa.int/web/gaia}, ASAS-SN\footnote{http://www.astronomy.ohio-state.edu/asassn/index.shtml}, Pan-STARSS\footnote{https://panstarrs.stsci.edu/}, iPTF \footnote{https://www.ptf.caltech.edu/iptf}) have significantly increased the number of EXor/FUor candidates, which suggests that episodic accretion is much more common behaviour for YSOs than previously thought. 

\begin{figure}[!t]
    \centering
    \includegraphics[width=6 cm]{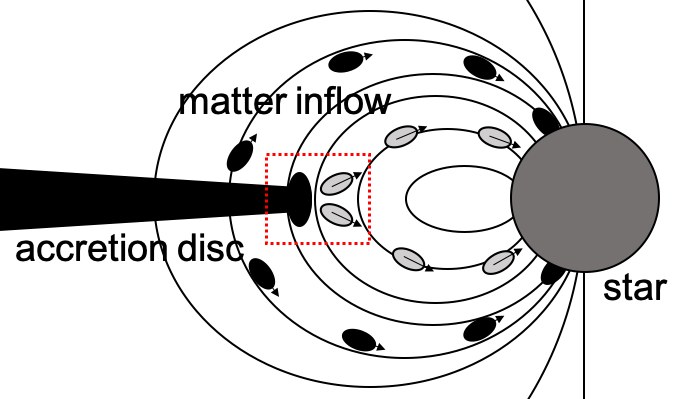}
    \caption{Cartoon of an accreting disc surrounding a young forming star. Lines of the idealized star dipolar magnetic field are represented as well, connecting the disc and star. Standard accretion inflows, following magnetic field lines, are represented as black droplets. Possible alternative inflow propagation across the magnetic field is represented as grey droplets.}
    \label{fig:accretion}
\end{figure}

In recent years, significant effort has been made to describe the process of accretion of mass in young stellar objects (YSOs) through the development of accurate multidimensional magnetohydrodynamic (MHD) models.  These theoretical and modelling studies have been very successful in describing the mass accretion in CTTSs and, in many cases, in compact objects as neutron stars as well. These studies showed that the matter, in proximity of the disc truncation radius, may flow along magnetic field lines of the magnetosphere (shown in Fig. \ref{fig:accretion}) connecting star and disc (\citealt{2008apsf.book.....H}).
Accretion would thus proceed as sketched by the dark drops in Fig. \ref{fig:accretion}, forming accretion funnels that hit the stellar surface at high latitudes (e.g.~\citealt{1990RvMA....3..234C}; \citealt{1991ApJ...370L..39K}; \citealt{2002ApJ...576L..53K}; \citealt{2002ApJ...578..420R}; \citealt{2003ApJ...595.1009R}; \citealt{2004ApJ...610..920R}; \citealt{refId0arg2017}). Another possibility is that accreting plasma may penetrate the magnetosphere in the equatorial plane, as sketched by the grey drops in Fig. \ref{fig:accretion}, through the development of Rayleigh-Taylor instability, forming thin tongues of plasma accreting onto the central protostar (e.g.~\citealt{1976ApJ...207..914A}; \citealt{1978ApJ...219..617S}; \citealt{2008MNRAS.386..673K}). The above possibility was numerically investigated in detail in \cite{2008MNRAS.386..673K}, however, there are still no sources based on astronomical observations supporting this hypothesis. Mass accretion can be even triggered by episodic perturbations of the accretion disc induced by an intense flaring activity occurring in proximity of the truncation radius (e.g.~\citealt{2011MNRAS.415.3380O}; \citealt{2017A&A...600A.105B}; \citealt{2019A&A...624A..50C}). All these models describe the complex dynamics of magnetospheric accretion through an almost continuous formation and disruption of accretion streams, which may account for the relatively short timescale (lasting from a few hours to several weeks) variability of mass accretion rates and luminosity observed in many YSOs (e.g.~\citealt{1996A&A...307..791G}; \citealt{ 1998ApJ...494..336S}; \citealt{2007A&A...463.1017B}; \citealt{2009ApJ...706..824C}; \citealt{2010A&A...519A..88A}; \citealt{Stauffer_2014}).

The above models, however, cannot account for the episodic outbursts with large amplitude and long timescales such as those observed in EXor and FUor systems. The physical origin of the sudden large increase in mass accretion rates in YSOs is still largely debated in the literature, mainly as a result of the paucity of observed events and their rarity. Possible explanations for the FUor/EXor outbursts have been proposed in the literature. The theoretical models explaining the origin of the outbursts can be roughly grouped in three main categories (\citealt{2019SAAS...45....1A}). A first class of models invokes a classical thermal instability (or secular instabilities; e.g.~\citealt{2001MNRAS.324..705A}) that occur in the accretion disc on a length scale of less than 1 AU (e.g.~\citealt{1994ApJ...427..987B}; see also \citealt{2001NewAR..45..449L} for the case of dwarf nova outbursts). According to this model, the star-disc system is unable to accrete at a steady rate (as, for instance, in the models discussed in the previous paragraph). As a result, the system alternates phases in which the gas gradually accumulates at the truncation radius, producing low accretion rates, to phases in which an instability (e.g. a thermal instability) may strongly perturb the disc and, possibly, trigger a high accretion event. A second class of models suggests that the episodic outbursts observed in EXor/FUor systems are triggered by a strong disc perturbation by an object external to the star-disc system.
The perturbation may be due to the gravitational effects of a binary companion (\citealt{1992ApJ...401L..31B}), or to the tidal effects by the migration of giant planets in the disc (\citealt{2004MNRAS.353..841L}; \citealt{2012MNRAS.426...70N}; \citealt{ 2005ApJ...633L.137V}).  In these cases, the disc can be affected by strong perturbations that trigger local instabilities (e.g. thermal instability) and lead to an event of high-accretion of mass. In a third class of models, the sporadic outbursts can be triggered by sudden changes in the stellar magnetic activity (\citealt{1995MNRAS.274.1242A}; \citealt{2016ApJ...833L..15A}; \citealt{2012MNRAS.420..416D}). Recent studies in the X-ray band have shown that the activity cycle of young stars can be highly irregular and have sudden changes in the activity level (e.g.~\citealt{2013A&A...553L...6S}; \citealt{ 2020A&A...636A..49C}); these changes are also characterized by high amplitude in the level of magnetic activity that may perturb the disc and trigger high-accretion events. However, despite significant theoretical and modelling efforts, to date none of the proposed models are able to provide a fully compelling description of the EXor/FUor phenomenon. For instance it is still not clear if EXors and FUors are different manifestations of the same phenomenon or if they represent two different classes of objects. We ask if they are triggered by the same physical mechanism and occur in the same evolution phase of a young accreting star.

From  observations (see also Table~\ref{tabEXor} below) we can gather the following: values of the density, n $\sim$ 10$^{12}$ - 10$^{14}$ cm$^{-3}$ are reasonable, using as a starting guess the value in standard accretion for CTTSs (see $5 \times 10^{11}$ cm$^{-3}$, as in \citealt{Bonito_2014}; \citealt{refId02009}).
For the temperature, T $\sim$ 10$^4$ - 10$^6$ K are reasonable values (e.g. ZCMa emits in X-rays, see also values for its jet in X-ray in \citealt{Kastner_2002}; \citealt{refId02007}; \citealt{bop10}; \citealt{bom10}).
The magnetic field of EXor-type objects is still unknown owing to the low statistics of observations for this class of objects. What is known is that the field in CTTSs, as measured on the surface of the stellar object, is, on average, of a few kG (\citealt{Johns_Krull_2007}; \citealt{Johns_Krull_2009}; \citealt{10.1111/j.1365-2966.2010.18069.x}; \citealt{10.1111/j.1365-2966.2011.19366.x}). Tens of G are thus reasonable values for the B-field strength at the distance corresponding to the disc truncation radius.

Our endeavour in this paper, since the parameters of that type of phenomena are poorly known, is hence to improve our knowledge of the parameters of the inflows that are at play. For this, we rely on laboratory experiments in which we produce different types of inflows propagating across magnetic field lines (as sketched in Fig.~\ref{fig:accretion}). The underlying idea is that, by showing that such inflow is possible, this gives weight to the possibility of having more accretion channels, and hence more mass accreting on the star, compared to having accretion relying solely on processes at high latitudes at which the plasma is guided by the magnetic field (\citealt{refId0arg2017}). This would thus be a way to provide more mass, but the initiation of this process would still rely on instabilities, as described above, which are not part of the laboratory experiments. In particular, in the present work, we do not discriminate between the conditions in which instabilities at the disc edge would occur. We merely show that plasma flows that can be scaled to what is known from EXor-type inflows can be measured and characterized well in the laboratory. 

We verify that these laboratory flows can be quantitatively scaled to those that are inferred from EXor observations, as detailed below and in Table~\ref{tabEXor}. Then, we use the detailed knowledge we have of the laboratory flows to infer parameters of the EXor inflows that are not accessible in the present state of the observational capabilities, among which is the magnetic field. 

Complementary to observations and numerical simulations of the processes, experiments conducted in the laboratory offer the possibility to gain further physical insight into the underlying physics on a wide range of topics, from astrochemistry (\citealt{munoz}) to plasma phenomena (\citealt{RevModPhys.78.755}). In contrast to computer modelling, experiments involve real phenomena, which are not based on approximations, assumptions, and idealizations as the models are. {}Experiments moreover include the full non-linearity of the processes. In contrast to observations, experiments conducted in the laboratory can in general be run and re-run many times, while varying some input parameters (e.g. the plasma temperature, ambient magnetic field, plasma density, and plasma composition) to assess their impact on the final observed plasma morphology and dynamics. However, the main and obvious difficulty is to assess how scalable, and hence relevant, such experiments are to astrophysical phenomena. Several studies have looked in detail at this issue of scalability. This is discussed in more detail in section \ref{sec:scal} below, but the general idea is to demonstrate that the main governing dimensional parameters of the laboratory and astrophysical plasmas are the same. This ensures that their dynamics is the same. The dimensional parameters allow us to retrieve the scaling in time, space, density, velocity, magnetic field, and so on between the two plasmas. Obviously, experiments are conducted on plasmas that have much smaller extent (mm to m) than the astrophysical plasmas, but the duration of the events are also much shorter durations (ns to s). This has the advantage that the laboratory plasma dynamics can be usually followed over a longer duration (when scaled) than what is accessible to the astrophysical observations. A constraint however is that, as there is a self-organisation regulating the magnetic field universally proportionally to the astrophysical system size (\citealt{doi:10.1146/annurev.aa.22.090184.002233}), usually the laboratory ambient magnetic field has to be quite strong (10-100 G to MG), so that it can have an effect on the plasma dynamics over the short spatial and temporal ranges at play.

Laboratory plasmas can be produced by a wide variety of means, for example discharges, plasma guns, pinches, or lasers. In a series of previous works (\citealt{2017SciA....3E0982R}; \citealt{HIGGINSON201748}; \citealt{burdonov2020laboratory}), we have already shown that, using high-power lasers coupled to strong external magnetic field, we could generate plasmas that scale to accretion funnels of CTTSs, those that follow magnetic field lines, and that give rise to the standard observed accretion rates.  

We explored the configuration in which plasma propagation takes place across the magnetic field lines, that is in a configuration that would be akin to accretion in the equatorial plane, as sketched by the grey droplets in Fig. \ref{fig:accretion}. The red dotted rectangle in Fig. \ref{fig:accretion} indicates the situation modelled in this article. We note that such a configuration has already been investigated in a number of earlier studies performed using various plasma machines (\citealt{doi:10.1063/1.1762404}; \citealt{doi:10.1063/1.1693157}; \citealt{PhysRevLett.111.185002}; \citealt{Tang_2018}; \citealt{khiar2019laser}). In this work, we prolong these previous efforts by generating plasmas that are scaled to the parameters of the massive accretion events which are our focus, as detailed below and in Table~\ref{tabEXor}. This is done using two different methods and facilities, namely a plasma gun and a high-power laser, the details of which are presented in section \ref{sec:lab.exp}. The use of two types of plasma-generating devices is mainly done to have two different sets of parameters (e.g. flow velocities, magnetic fields, and plasma densities), helping to constrain the parameters that could be at play in EXors. 

We find that, first, consistent with previous studies, in both cases efficient plasma propagation across the magnetic field lines is possible, thus giving weight to the hypothesis of massive accretion through the disc gap that was proposed by numerical simulations. Second, the laboratory streams parameters can be scaled to those giving rise to the massive outbursts detailed above, using the estimated parameters that can be retrieved from the few observations we have of the latter. Doing this, an interesting point is that the scaling of the laboratory magnetic field gives a hint of the magnetic field that would be at play in the astrophysical situation. As magnetic fields have not been primarily derived from observations up to now, the laboratory experiments are thus not only grounding the possibility of massive accretion across magnetic field, as shown in Fig. \ref{fig:accretion}, but further indicate in which range of magnetic field this is possible. 

The paper is organized as follows. In Sect. \ref{sec:lab.exp}, we describe the facilities used for the laboratory modelling; in Sect. \ref{sec:lab.result}, we present the experimental results complemented by numerical simulations; in Sect. \ref{sec:scal}, we discuss the scalability of the laboratory experiment to the astrophysical phenomena of interest and compare the laboratory parameters with the particular object representing the episodic accretion event; and in Sect. \ref{sec:conc}, we discuss the results and draw our conclusions regarding the magnetic field at play in the scaled EXor situations.

\section{Laboratory experimental approach}
\label{sec:lab.exp}

In this section we present the two different facilities (plasma gun and high-power laser), which we used to investigate the dynamics of plasma propagation across a strong external magnetic field. In both cases, as illustrated in Fig. \ref{fig:setup}, we have a hot plasma that is generated either by the plasma gun or ablated by the high-power laser from a solid target, which expands in a vacuum. The whole configuration is embedded in a magnetic field (aligned with the x-axis) that is perpendicular to the main axis of the plasma expansion (z-axis). As illustrated by Fig. \ref{fig:setup}, as a consequence of the interaction between the hot, expanding plasma and the magnetic field, the plasma morphology, which is initially conical and has an opening angle around  40$^\circ$ for the plasma gun and around 30$^\circ$ (\citealt{revet:tel-02100492}) for the laser, becomes flattened along the y-axis (i.e. that perpendicular to the magnetic field) and stretched along the x-axis (i.e. that of the magnetic field). As shown in Sect. \ref{sec:lab.result}, this plasma structure propagates unimpeded across the magnetic field lines (i.e. in the direction of the z-axis).

\subsection{Using the high-power laser facility PEARL}
\label{sec:PEARL}

\begin{figure}[htp]
    \centering
    \includegraphics[width=6 cm]{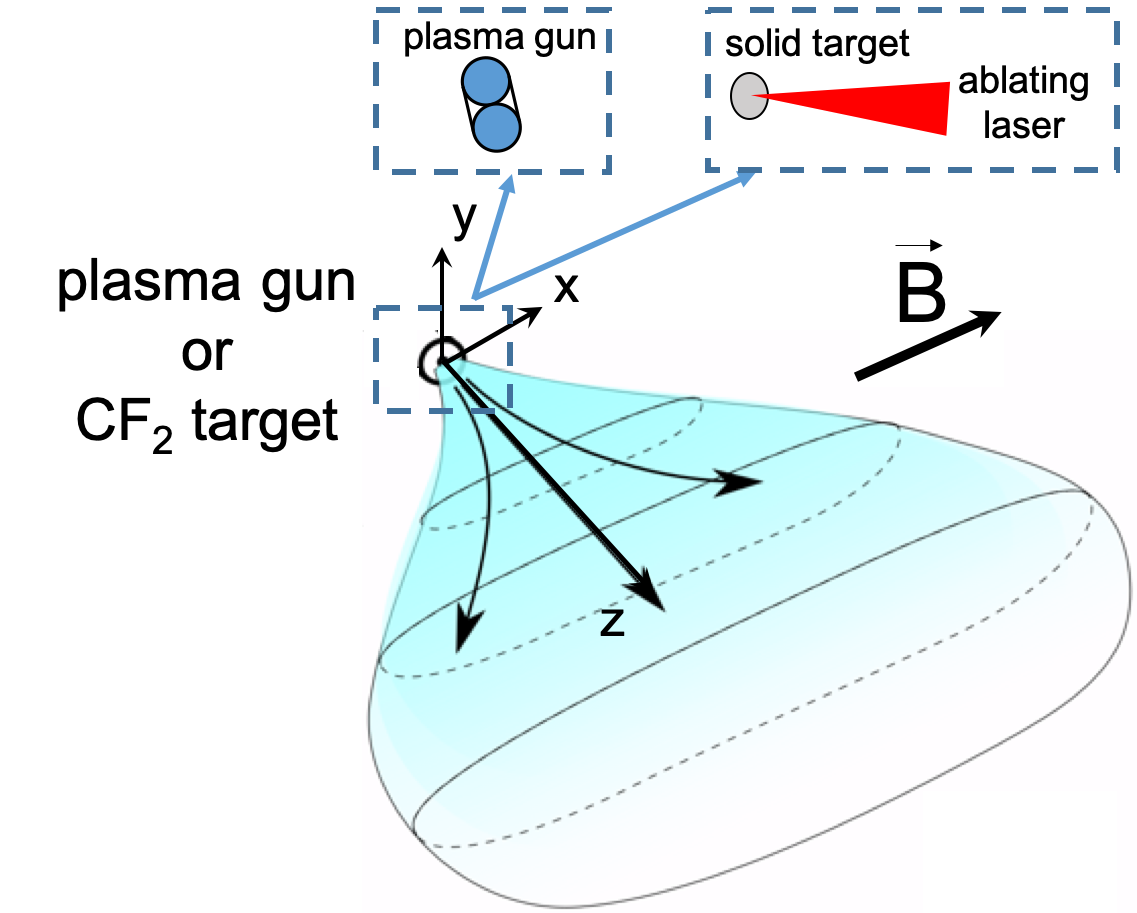}
    \caption{Schematic view of the experimental set-up. A plasma is generated from a plasma gun or a laser-ablated Teflon (CF$_{2}$) target. It expands into the vacuum (along z as the main expansion axis), which is embedded in a strong external magnetic field, orientated along x. Because of the imposed magnetization, the plasma is progressively flattened out along y, while it can flow along the magnetic field line (i.e. along x). Hence, it progressively morphs into a `pancake', as depicted.
    }
    \label{fig:setup}
\end{figure}

The high-power laser that is used to generate the expanding plasma is atthe PEARL laser facility (\citealt{Lozhkarev_2007}; \citealt{Soloviev2017}; \citealt{Perevalov_2020}) located at the Institute of Applied Physics of the Russian Academy of Sciences (IAP RAS), Nizhny Novgorod (Russia). The set-up is presented in the Figure \ref{fig:setup}. The laser pulse (3 J, 1 ns, 527 nm, normal incidence) irradiated the surface of a thick, solid CF$_{2}$ target that has its normal orientated along the z-axis. The laser, ablating the surface target material, induced the expansion of a plasma stream into the vacuum, along the z-axis and perpendicular to the externally imposed B-field of $1.35\times10^5$ G). The magnetic field was created by a Helmholtz coil that maintained a spatially ($\sim$ 2 cm) and temporally ($\sim$ 1 $\mu$s) constant field over the scales of the experiment (1 cm and 100 ns). The initial laser beam had a flat-top circular-shaped spatial profile with 100 mm diameter. Before being focussed, the beam was partially blocked by the opaque screen that has a 10 mm opening gap centred with the beam axis. As a consequence, the spot size on the target surface had a quasi-rectangular shape (0.1 cm by 1 cm) delivering a laser intensity around $3 \times 10^{10}$ W/cm$^{2}$.
Optical interferometry imaging using a Mach-Zehnder scheme was used to diagnose the plasma stream propagation. The optical probing used a low-energy (100 mJ), short-pulse beam (100 fs) to obtain a snapshot of the plasma as it evolved. This was made simultaneously along two probing axes, namely y and z, to analyse the 3D nature of the plasma flow. Such snapshots were captured at various times, up to 108 ns with 10 ns step, relative to the laser-ablating laser impacting the target.

\subsection{Using a plasma gun at the KROT facility}
\label{sec:KROT}

The large-scale (volume of 170 m$^{3}$) KROT device (\citealt{aidakina2018simulation}) is the space plasma simulation chamber designed and constructed at the IAP RAS, Nizhny Novgorod (Russia). The chamber is equipped with a pulsed solenoid for the magnetic field generation and a pulsed radio frequency (RF) source of inductively coupled background plasma. The background plasma can be composed of argon or helium, and has a maximum density of n$_{e}$ $\sim$ 10$^{12}$ - 10$^{13}$ cm$^{-3}$; the electron temperature reaches several eV. The magnetic field is set in a mirror trap configuration, has a trap ratio R = 2.3, and a strength up to 450 G in the central cross section. Around the centre of the trap, the area with quasi-uniform field (i.e. it varies less than 10\% within this region) has a length of about 1 m and diameter not less than 1.3 m. The large volume of uniform magnetic field along and across field lines is a unique feature of the KROT plasma chamber.

The plasma flow was injected into the vacuum at the residual air pressure p$_{0}$ $\sim$ 3 - 50 $\mu$ Torr and propagated in the ambient B-field having strength from 45 to 450 G. The overall set-up, illustrated in Figure \ref{fig:setup}, is similar to that of the high-power laser experiment. The plasma was generated by a "cable gun" made of a 50 Ohm coaxial cable with a polyethylene insulator (\citealt{gushchin2018laboratory}; \citealt{2019TePhL..45..228K}).
The gun was installed at the centre of the area of the quasi-uniform magnetic field. The plasma flow was induced via high-voltage surface breakdown on the insulator, with subsequent plasma acceleration by the Ampere force (\citealt{Marshall_1960}). The outer diameter of the gun was 10\,mm and its operating voltages and currents were around 5\,kV and 4\,kA, respectively. The current pulse duration was about 15 $\mu$s at its base, while the current rise time was 7 $\mu$s.

Several diagnostics were used to measure the plasma parameters \citealt{gushchin2018laboratory}. The self-emission from the plasma stream was recorded, using a 4Picos fast shutter camera, along two directions, parallel and perpendicular to the B-field lines axis. The snapshots were taken from 0 $\mu$s to 30 $\mu$s after the gun discharge ignition with 1 $\mu$s step. The local plasma density n$_{e}$ and electron temperature T$_{e}$ values were measured using double electric probes immersed into the plasma stream. The plasma density, averaged over the stream cross section, was measured using a microwave interferometer with an operating frequency of $37.5$\,GHz and from the cut-off of a probing microwave signal with the same frequency. Local magnetic disturbances were measured by a set of identical B-dot probes. The diamagnetic effect in the stream moving into the ambient magnetic field was used as an independent method for diagnosing the plasma parameters\ (i.e. its density and temperature).

\section{Laboratory plasma measurements}
\label{sec:lab.result}

In this section we present the results of the laboratory experiments performed as presented in the previous section, complemented by numerical modelling. Despite different spatial and temporal scales, in both cases the plasma flow demonstrated similar propagating features. The values of the plasma flow velocities, densities, and temperatures, presented in this section, are used in the following section to link the laboratory plasma with the astrophysical phenomena of interest.

\subsection{Measurements of the high-power laser generated plasma density and velocity}
\label{sec:lab.result.PEARL}

\begin{figure}[htp]
    \centering
    \includegraphics[width=9 cm]{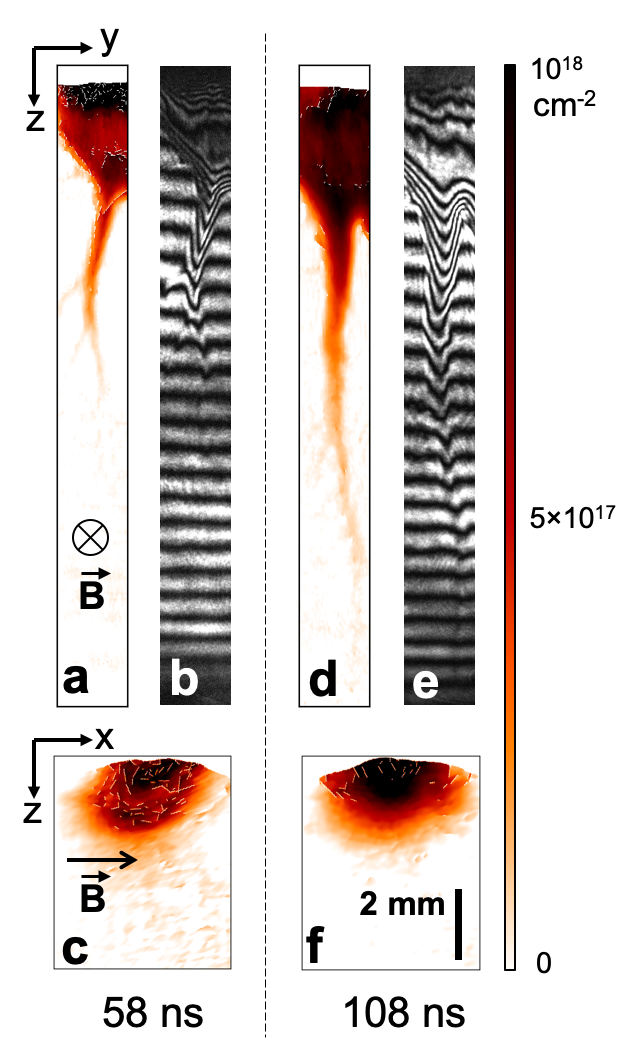}
    \caption{Line-integrated, 2D projections of the electron density as measured, 58 ns after the laser impact onto the target, (a) in the yz and (c) xz-planes. Panel (b) shows a raw interferometry image for the yz-plane for better illustration. Panels (d-f) shows the same at 108 ns after the laser impact onto the target. The projections in the two orthogonal planes are measured simultaneously.}
    \label{fig:stream.pearl}
\end{figure}

Figure \ref{fig:stream.pearl} shows the measurements of the laser-generated plasma flow. The fact that we irradiated a large surface of the target induced a complex, fragmented morphology of the plasma, similar to what has already been observed in current-driven plasma flows from thick rods \citealt{ivanov2006}. In this work we do not focus on characterizing the plasma structure, but rather focus on the global parameter (velocity, density, and temperature) of the central part of the expanding plasma. From Figure \ref{fig:stream.pearl}, we verify that the global morphology of the plasma is similar to that illustrated in Fig. \ref{fig:setup}; that is, the plasma can spread in the x-axis, while it is pinched along the y-axis, resulting in an overall pancake shape. The electron density value of the propagating flow at the distance of about 5 mm from the target surface was of the order of $5\times 10^{17}$ cm$^{-3}$. 

\begin{figure}[htp]    
\centering
\includegraphics[width=6 cm]{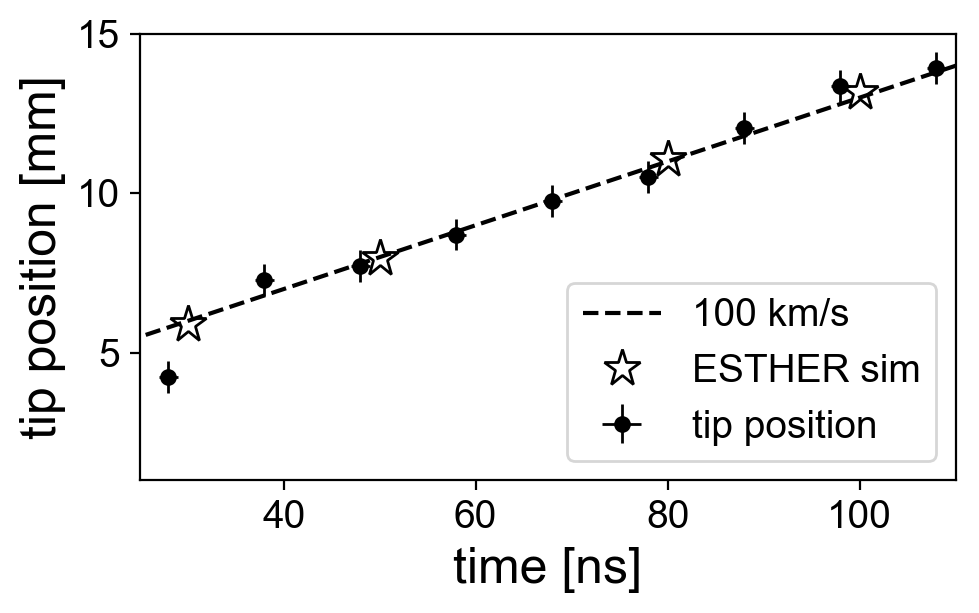}
\caption{Temporal evolution of the tip of the laser-generated plasma flow, as retrieved from images as those shown in Fig. \ref{fig:stream.pearl} (full dots) and as simulated by the ESTHER  code (empty stars). The dashed line corresponds to an average velocity of 100 km/s.}
\label{fig:tip_char}
\end{figure}

The temporal evolution of the tip of the plasma flow is shown in Figure \ref{fig:tip_char}. The tip position was defined as corresponding to 5$\times$10$^{16}$ cm$^{-2}$ in the 2D projection of the electron density in the yz-plane. Since the width of the plasma pancake (in the xz plane) is around 0.3 cm, that threshold thus corresponds to a volumetric density around 10$^{17}$ cm$^{-3}$. We observe that the tip of the plasma has a constant velocity of around 100 km/s.

\subsection{Simulations of the high-power laser generated plasma temperature}
\label{sec:ESTHER}

\begin{figure}[htp]    
\centering
\includegraphics[width=6 cm]{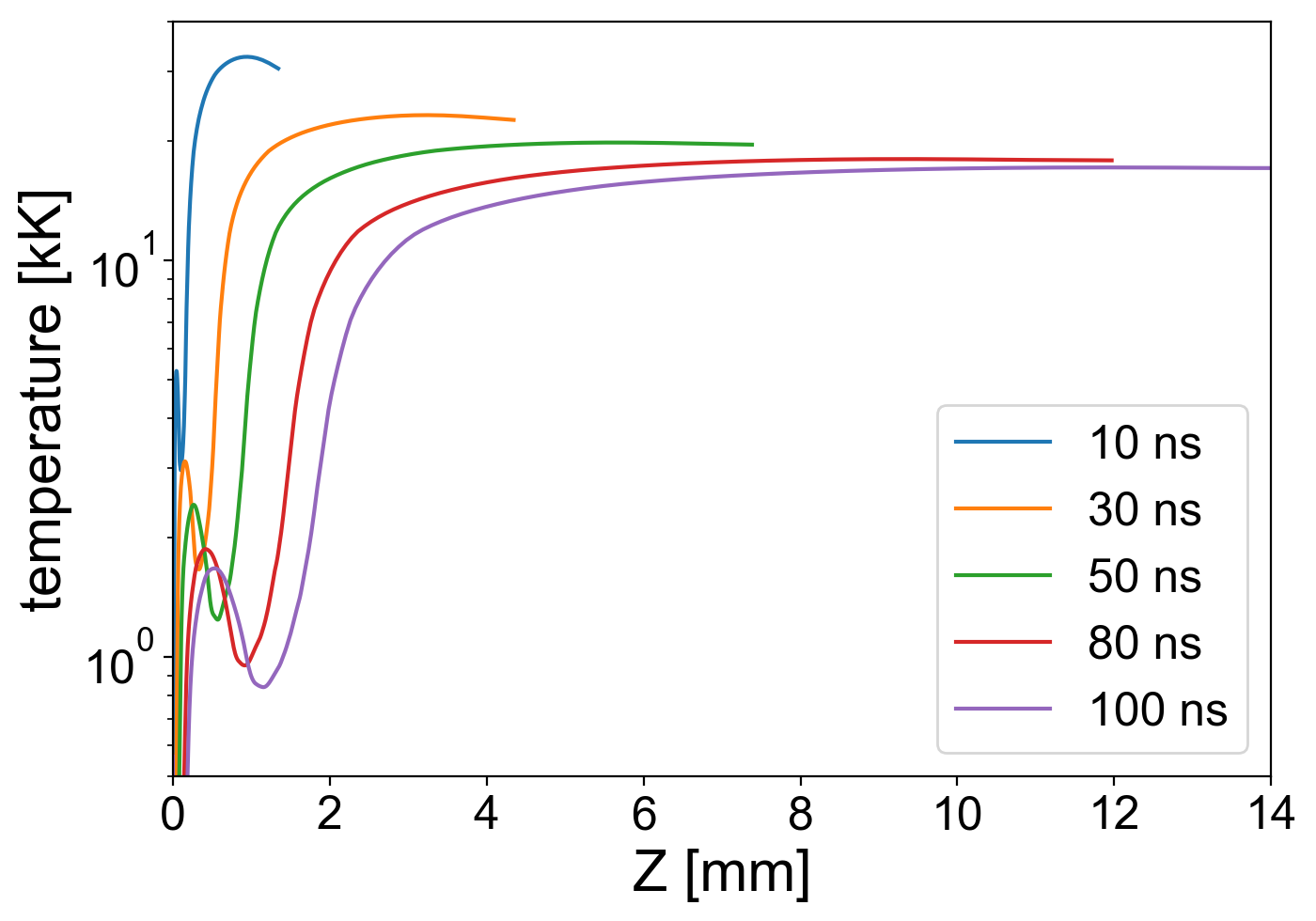}
\caption{Temperature distribution, over time, of a laser-generated plasma flow, as simulated with the ESTHER code, and using similar parameters as the experimental parameters (see text).}
\label{fig:temp_pearl}
\end{figure}

Since we could not directly measure the temperature of the plasma generated by the high-power laser, we rely on numerical simulations to retrieve the temperature of the propagating plasma pancake. For this, we use the Lagrangian 1D hydrodynamic code ESTHER (\citealt{PhysRevB.71.165406}). 
As shown schematically in Fig.2 (and experimentally in Fig.3 for the laser-produced plasma and Fig.6 for the plasma gun), the plasma flow is constrained in the shape of a pancake by the external magnetic field, that is it is flattened in the yz plane, but expands laterally in the xz plane. So, in the xz plane, the situation is 1D since the plasma is constrained by the B-field. In the xz plane, the situation is 2D, but since the radial size of the pancake expansion front is large, locally, the situation can be approximated in 1D. Hence, we simulate a central line of the expanding plasma with ESTHER. The simulation does not treat the background magnetic field; the justification for this is that within its expansion, as it propagates the plasma expels the magnetic field (\citealt{khiar2019laser}). This code solves, according to a Lagrangian scheme, the fluid equations for the conservation of mass, momentum, and energy. The target material is described by the Bushman-Lomonosov-Fortov (BLF) multi-phase equation of state spanning a large range of density and temperature from hot plasma to cold condensed matter. ESTHER is a single-temperature code, i.e. T$_e$ = T$_i$. We use this code since it specifically allows us to simulate the transition to plasma of a solid under the impact of a low intensity laser (\citealt{PhysRevB.71.165406}), which has an intensity of the order of 10$^{10}$ W/cm$^2$. Since we cannot simulate a composite material as that used in the experiment, a pure carbon target was simulated. Fig. \ref{fig:temp_pearl} represents the plasma temperature distribution from the target surface up to tens mm along the main expansion axis (the z-axis, see Fig. \ref{fig:setup}) at specific times 10 ns, 30 ns, 50 ns, 80 ns, and 100 ns after the impact of a laser pulse with $3 \times 10^{10}$ W/cm$^2$, which has a Gaussian temporal shape and a 1 ns FWHM duration. We observe that the plasma stabilizes around a temperature of $\sim$ 17 - 20 kK far from the target (i.e. at distances of $\sim$ 6 - 12 mm and times of $\sim$ 50 - 100 ns). The fact that the temperature is high only at the head of the expanding plasma is related to the fact that the laser energy is low, and hence only a thin layer on the target surface can be ablated and be at high temperature. The position of the tip in the simulations, presented in the Fig. ~\ref{fig:tip_char}, defined at the distance from the target surface, where the electron density equals to $1 \times 10^{17}$ cm$^{-3}$.

\subsection{Measurement of the plasma gun generated plasma}
\label{sec:lab.result.KROT}

\begin{figure}[htp]
    \centering
    \includegraphics[width=7 cm]{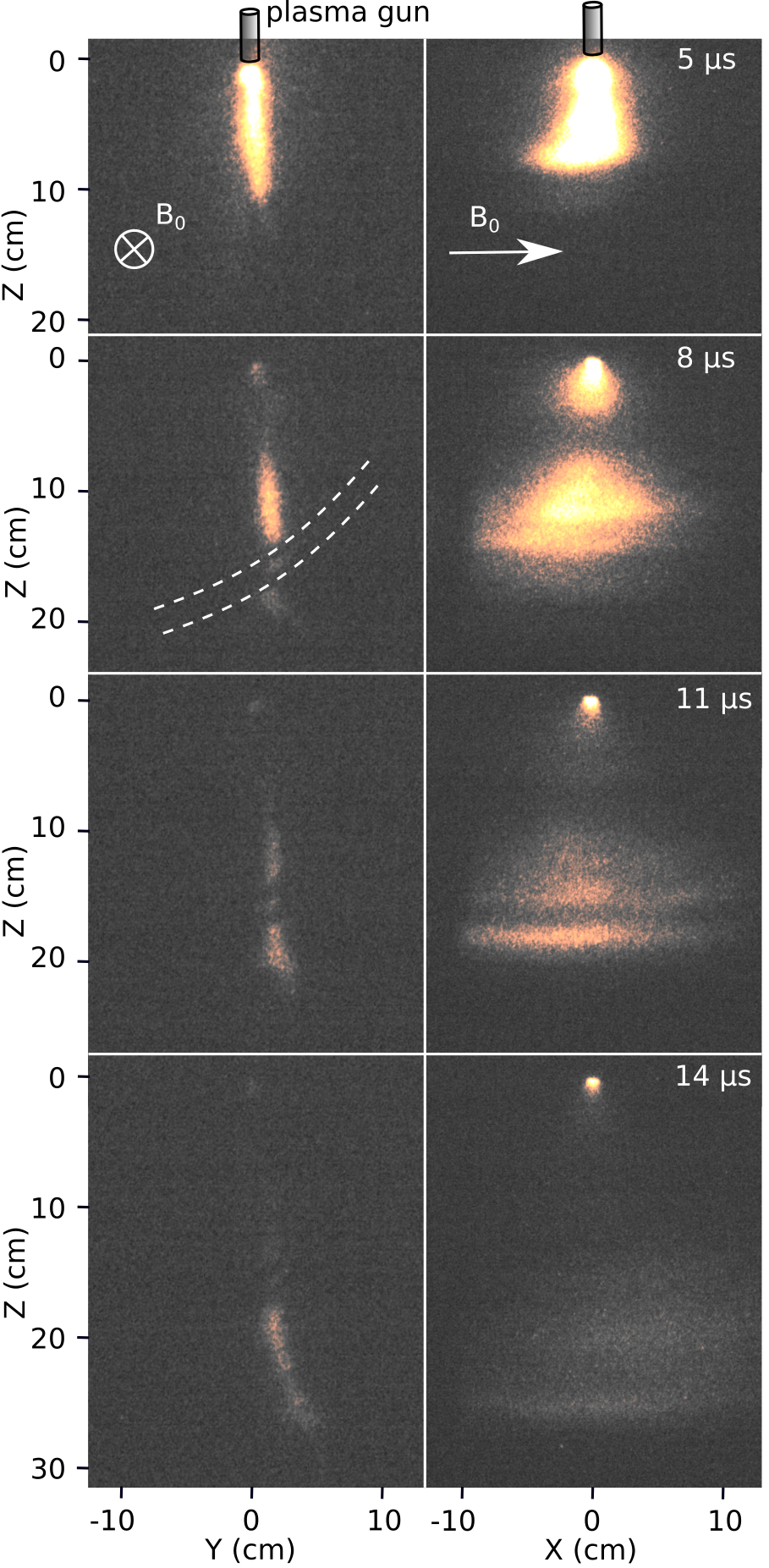}
    \caption{Images of the self-emission of the plasma-gun generated flow (in the case in which the applied magnetic field strength is 450 G), at various times after the start of injection (as indicated) and in the yz (left column) and xz (right column) planes.  The dashed lines denote the location of a cable line placed between the plasma flow and the observation camera, hence locally obscuring the flow. }
    \label{fig:stream.krot}
\end{figure}

\begin{figure}[htp]
    \centering
    \includegraphics[width=7 cm]{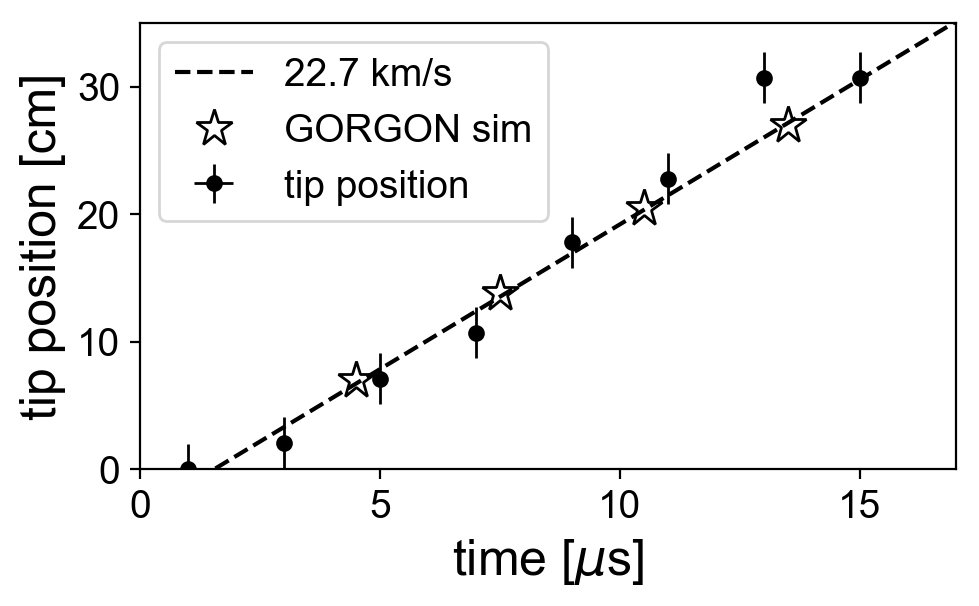}
    \caption{Temporal evolution of the tip of the plasma gun-generated flow (in the case in which the applied magnetic field strength is 450 G), as retrieved from images as those shown in Fig. \ref{fig:stream.krot} (full dots) and as simulated by the GORGON  code (empty stars). The dashed line corresponds to an average velocity of 22.7 km/s.}
    \label{fig:tip.krot}
\end{figure}

As can be seen in Fig. \ref{fig:stream.krot}, the plasma gun-generated flow demonstrates the same topological features as the laser-generated flow. Without a magnetic field, the plasma expanded into a cone that has an opening angle of (37 $\pm$ 3)$^\circ$. In the presence of the magnetic field, the plasma flow was transformed into a pancake structure, similar as that of the laser-generated plasma (compare Fig. \ref{fig:stream.pearl} and Fig. \ref{fig:stream.krot}). For magnetic field strengths higher than 100 G, the pancake thickness was inversely proportional to the magnetic field strength. For the maximum magnetic field, B$_{0}$ = 450 G, the pancake thickness was about $\delta$y = 2 cm. This pancake-shaped plasma flow propagates with its main axis along z (see Fig. \ref{fig:setup}), that is perpendicularly to the average direction of the magnetic field, up to 50 - 70 cm from the injection point and with almost constant velocity (around 23 $\pm$ 1) km/s (see Fig. \ref{fig:tip.krot}). The tip position was defined as the position at which the plasma luminosity was at the level of $\sim$ 0.25 from the averaged, which corresponds to the electron plasma density of the order of several $10^{12}$ cm$^{-3}$. Additional features of the plasma flow dynamics in a transverse magnetic field were the presence of a twist of the stream in the direction of ion gyration (see Fig. \ref{fig:stream.krot}, on the left) and the development of an observed transverse fragmentation of the flow (see Fig. \ref{fig:stream.krot}, on the right).

The measured electron density was n$_e$ $\sim$ 10$^{13}$ cm$^{-3}$ inside the pancake and above 10$^{14}$ cm$^{-3}$ near to the injection point. The electron temperature was up to $\sim$ 46 kK during the gun discharge at the time of maximum current (around 7 $\mu$s after the start of injection). Later, at the stage of the pancake plasma free expansion, the electron temperature was around 12 kK. The level of diamagnetic disturbances inside the pancake was not higher than several G and could be attributed to the thermal diamagnetism of the plasma. 

The presence of an optional background argon plasma, which has a density of up to 10$^{12}$ cm$^{-3}$ and temperature T$_{e}$ $\sim$ 12 kK, affected neither the observed dynamics of plasma flow nor the measured plasma parameters for the same range of magnetic field strengths. The formation of the observed pancake structure was a persistent physical effect, which was insensitive to the presence of a background plasma with an electron density comparable to that of the plasma flow. 

\subsection{Simulations of the plasma gun generated flow}
\label{sec:GORGON}

We used the resistive, single-fluid, bi-temperature and highly parallel code GORGON (\citealt{ciardi2007evolution}; \citealt{ciardi2013astrophysics}) to perform 3D MHD simulations of the plasma gun generated flow. Since the laboratory measurements are performed with probes inserted in the flow, the simulations can be used as a check that the measured parameters match those of the (simulated) unperturbed flow.

The simulation box is defined by a uniform Cartesian grid of dimension $20 \times 20 \times 30$ cm$^3$ and the number of cells equals to $160 \times 160 \times 240$. The spatial resolution is homogeneous, its value is $d_x = d_y = d_z = 1.25$ mm, and the simulation lasts for 30\,$\mu$s.

The plasma (CH$_2$) is injected from the centre of the boundary at $z=0$ with the same diameter as the KROT injector. The initial density is $3\times 10^{13}$ cm$^{-3}$, the velocity is 25 km/s (with 40 $^\circ$ cone angle), and the temperature is $\sim$ 46 kK. A uniform B-field along the x-direction is applied, with a series of strength (i.e. $B_x = 56/112/225/450$ G) and cases with and without background cold helium gas with a number density of $10^{12}$ cm$^{-3}$ are compared.

\begin{figure}
    \centering
    \includegraphics[width=9 cm]{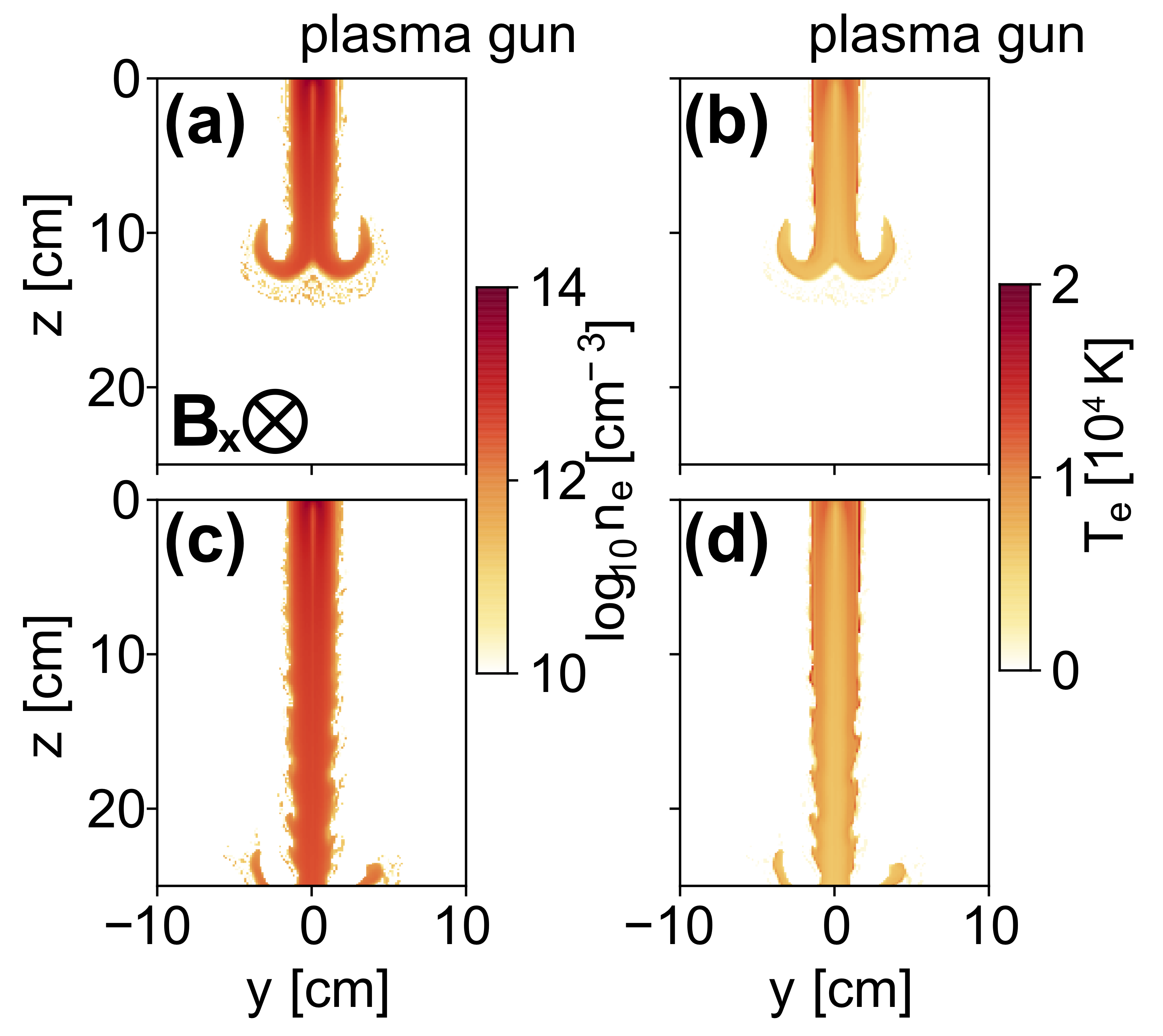}    \caption{Three-dimensional MHD simulations of the plasma gun generated flow performed with the GORGON code. Two-dimensional distributions of the plasma in yz-plane, passing through the middle of the 3D flow, are shown. Panels (a) and (b) are at t = 6 $\mu$s, while panels (c) and (d) are at t = 12 $\mu$s. The plasma gun is at the top, like in the experiment (see Fig. \ref{fig:stream.krot}). The electron number densities [cm$^{-3}$] are shown on the left in logarithmic scale, while the electron temperatures [Kelvins] are on the right. The applied magnetic field is B$_x$ = 450 G, directed out of plane.}
    \label{fig:KROT_GORGON}
\end{figure}

The simulation results for $B_x = 450$ G are shown in Fig.~\ref{fig:KROT_GORGON}. Electron number densities are on the left, while electron temperatures are on the right. The tongue structures seen at the tip of the plasma jet in Fig.8 originate from the initialization of the simulation, namely from how we inject plasma from the boundary of the simulation box. These structures are not important for plasma dynamics and we do not observe well-developed Rayleigh-Taylor instability or the Kelvin-Helmholtz instability within the timescale of interest in our simulations.

First and foremost, we observe that the GORGON simulation reproduces the pancake structure from the KROT experiment and has a stable width of $\sim 3$ cm. The results in the xz-plane merely display a homogeneous plasma in density and temperature and are thus not shown in this work.
Besides, the simulation confirms the characterization of the plasma parameter from the experiment: the electron density of the stream is around $10^{13}$ cm$^{-3}$, the electron temperature is around 12 kK, and the velocity of the plasma flow is more than 20 km/s.
The criteria to define the tip position, presented in the Fig. ~\ref{fig:tip.krot}, is $2.0 \times 10^{12}$ cm$^{-3}$ in the yz-plane sliced at x = 0.

Moreover, by comparing the cases with and without background gas (not shown here), we see that the pancake structure is almost the same, confirming that the effect of a background can be neglected, as found in the plasma gun experiments.

\section{Scaling of the laboratory plasmas to EXor objects}
\label{sec:scal}

Now that we detailed the parameters of the laboratory propagating plasmas, we describe how they can be quantitatively scaled to those of the EXors accreting inflows. This section describes that scalability approach used for the comparison of the laboratory plasma with the astrophysical plasma. For the latter, we focus on the estimated parameters of the incoming stream before its impact on the surface of the star, as derived in \cite{Giannini_2017}.

\subsection{Principle of the scalability}
\label{sec:scal_descr}

The scalability between the laboratory plasma and the astrophysical plasma is based on the approach proposed in the works by Ryutov (\citealt{Ryutov_1999}; \citealt{Ryutov_2000}; \citealt{doi:10.1063/1.5042254}).
The "Euler similarity" is based on two scaling quantities: the Euler number (Eu = V($\rho$/p)$^{1/2}$, here and below the formulas are presented for cgs units) and the plasma beta ($\beta$ = 8$\pi$ p/B$^{2}$), where V is the flow velocity, $\rho$ is the mass density, B is the magnetic field, and p = k$_{B}$(n$_i$T$_{i}$+n$_{e}$T$_{e}$) is the thermal pressure; k$_{B}$ is the Boltzmann constant and n$_{i,e}$ and T$_{i,e}$ are the number densities and temperatures of the ions and electrons, respectively. When matching for two different systems, it can be shown that these two systems are scaled to each other and evolve identically.

The Euler similarity holds and allows us to use ideal MHD equations if dissipative processes, which might affect the fluid dynamics, are neglected. The Reynolds number (the ratio of the inertial force to the viscous force) is responsible for the viscous dissipation, the magnetic Reynolds number (the ratio of the convection over Ohmic dissipation) for the resistive diffusion, and the Peclet number (the ratio of heat convection to the heat conduction) for the thermal conduction. All these parameters should be higher than 1 to meet the required conditions.

We note that for the episodic accretion EXor events, not all the parameters necessary for the comparison with the laboratory systems are known. The analysis of the astronomical observations allows us to estimate the plasma densities, temperatures, and characteristic velocities of the streams. However, as mentioned in Sect.~\ref{sec:intro}, the magnetic field of the involved protostars is not known. Some works report estimations of kilogauss levels (\citealt{donati2005direct} (FU Ori); \citealt{green2013variability} (HBC 722)) while for the EXors the magnetic field is supposed to be weaker (\citealt{2014prpl.conf..387A}), of the order of hundred G. One of the objectives of the present work is to use the scalability to the laboratory flows to estimate the possible values of the B-field strength of the EXor objects.

\subsection{Comparison of the laboratory flows to the accretion events of EXor objects}
\label{sec:EXor}

In the paper \citealt{Giannini_2017}, the parameters of the observed episodic accretion event of a particular EXor object were presented. The typical densities of the accretion stream in the rising/peak phase  range from $4 \times 10^{11}$ cm$^{-3}$ to $6 \times 10^{11}$ cm$^{-3}$, the temperatures range from 9 kK to 15 kK and the stream velocities range from tens to hundreds km/s. But no information about the B-field strength can be retrieved from such observations.

\begin{figure}[htp]    
\centering
\includegraphics[width=8 cm]{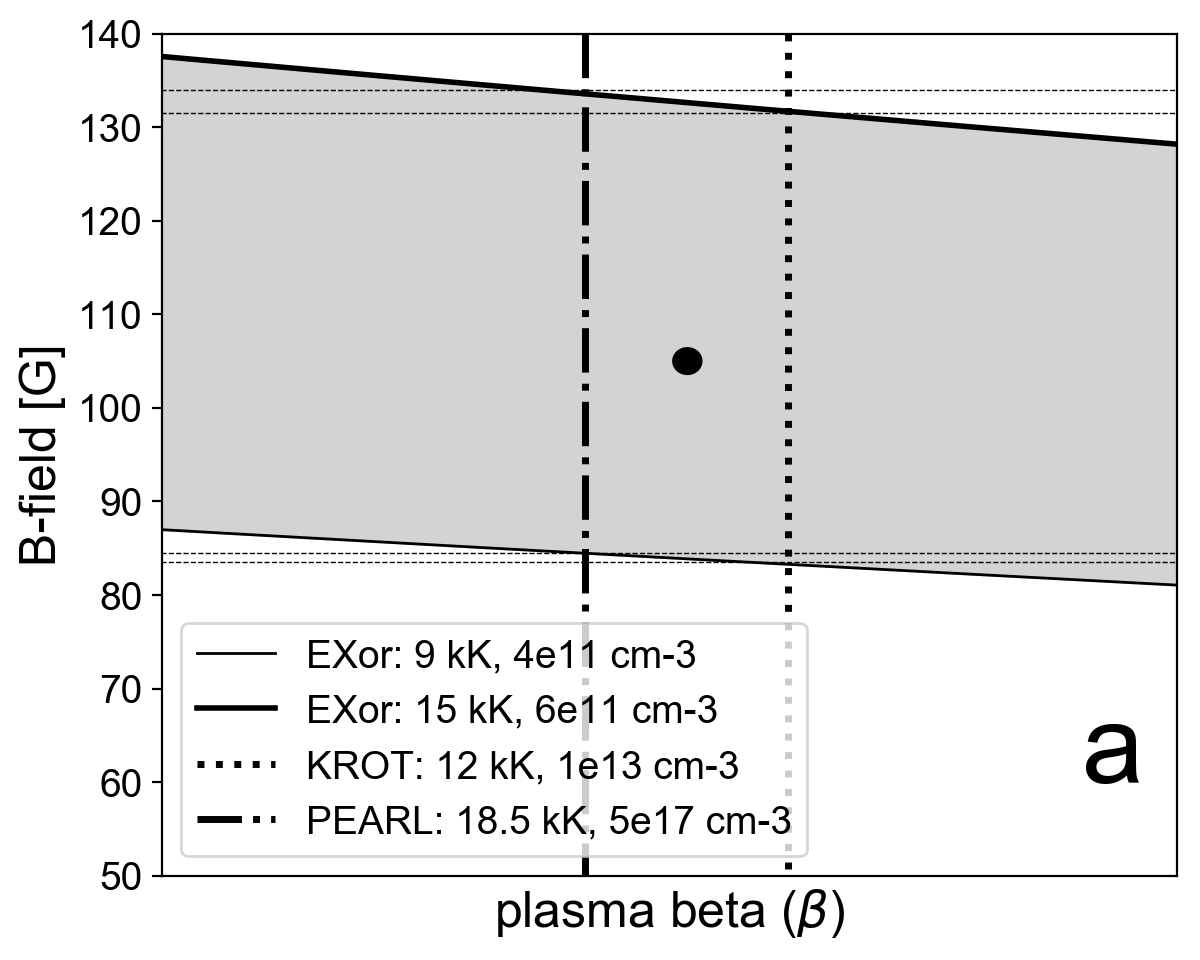}
\includegraphics[width=8 cm]{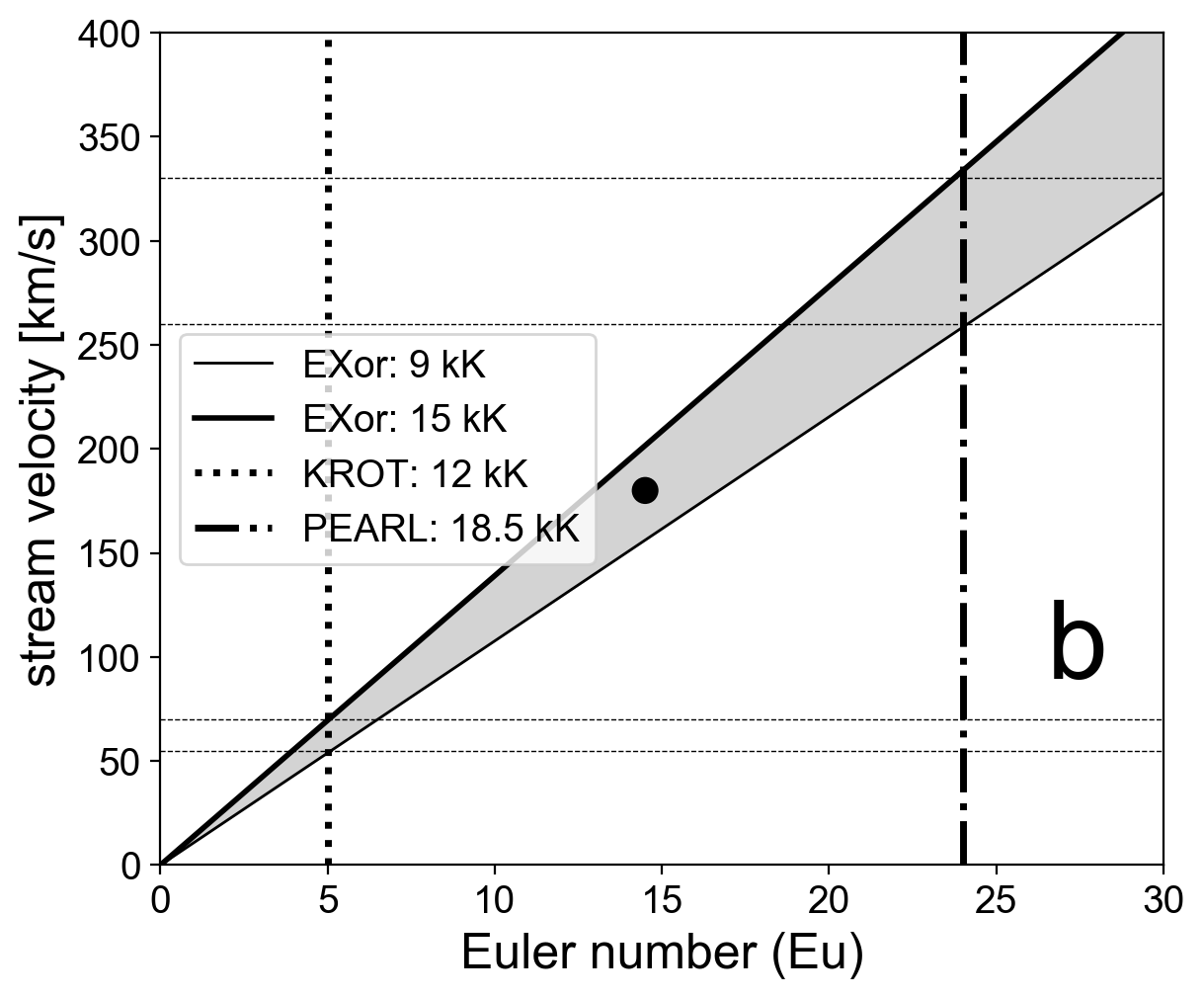}
\caption{(a) Correlation between the magnetic field and the $\beta$. The grey area represents the range of EXor parameters, the dot-dashed vertical line indicates the laser-driven flow $\beta$, the dotted line indicates the plasma gun-driven flow $\beta$, and the black point represents the parameters of the EXor used in the Table~\ref{tabEXor}. (b) Same for the correlation between the stream velocity and the Euler number.}
\label{fig:EXor_beta_Eu}
\end{figure}

Figure \ref{fig:EXor_beta_Eu} represents the phase-space of the possible plasma beta and Euler number of the considered EXor object, depending on the B-field strength and the stream velocity propagation, respectively. The grey areas correspond to the estimated range of parameters for accreting inflows in EXor objects. The dot-dashed vertical lines denote the plasma beta corresponding to the laser-driven flow, while the dotted lines correspond to the plasma gun-driven flow. From the Figure \ref{fig:EXor_beta_Eu} (a) we see that both the laser-driven flow and the plasma gun-driven flow scale to the EXor B-field in the range from $\sim$ 85 to $\sim$ 135 G. Figure \ref{fig:EXor_beta_Eu} (b) shows that the EXor accreting inflow velocity ranging from 260 to 330 km/s scales to the laser-driven flow and from 55 to 70 km/s to the plasma gun-driven flow. The black points in the figure indicate the typical averaged parameters of EXor used in  Table~\ref{tabEXor}; B-field is assumed to be 105 G.

In the comparative Table~\ref{tabEXor}, the parameters of the laser and plasma gun-driven laboratory streams and the considered EXor accretion stream (denoted with the black point in Figure \ref{fig:EXor_beta_Eu}) are listed. All parameters are calculated for plasma inside the propagating stream; no shocks are considerable. The symbols in Table~\ref{tabEXor} are as follows: Z is the charge state; A is the mass number; B is the magnetic field; L is the  spatial scale; n$_{e}$ is the electron density; $\rho$ is the mass density; T$_{e}$ is the electron temperature; T$_{i}$ is the ion temperature; V$_{flow}$ is the stream velocity; C$_{S}$ is the sound velocity (C$_s$ = ($\gamma$k$_{B}$(n$_i$T$_{i}$+n$_{e}$T$_{e}$)/$\rho$)$^{1/2}$, where $\gamma$ = 5/3 is the adiabatic index); V$_{A}$ is the Alfven velocity (V$_{A}$ = B/$\sqrt{4\pi n_i m_i}$, where $n_i$ is the ion density and $m_i$ is the ion mass); $\tau_{col\:e}$ is the electron collision time (\citealt{Braginskii_1965} at p. 215); l$_{e}$ = V$_{th\:e}\tau_{col\:e}$ is the collisional electron mean free path, where V$_{th\:e}$ is the electron thermal velocity (\citealt{NRL_pf} p. 29); R$_{Le}$ is the electron Larmor radius, f$_{ce}$ is the electron gyrofrequency; $\tau_{col\:i}$ is the ion collision time (\citealt{Braginskii_1965} at p. 215); l$_{i}$  = V$_{th\:i}\tau_{col\:i}$ is the collisional ion mean free path, where V$_{th\:e}$ is the ion thermal velocity (\citealt{NRL_pf} p. 29); R$_{Li}$ is the ion Larmor radius, f$_{ci}$ is the ion gyrofrequency; $M$ = V$_{flow}$/C$_s$ is the Mach number; M$_{alf}$ is the Alfven Mach number (M$_{alf}$ = V/V$_{A}$); $\tau_{\eta}$ is the magnetic diffusion time ($\tau_{\eta}$ = L$^{2}/\eta$, where $\eta$ is the magnetic diffusivity (\citealt{Ryutov_2000} p. 467); Re$_{M}$ is the magnetic Reynolds number (Re$_{M}$ = $LV/\eta$); Re is the Reynolds number (Re = LV/$\nu$, where $\nu$ is the kinematic viscosity (\citealt{Ryutov_1999} p. 826); Pe is the Peclet number (Pe = LV/$\chi$, where $\chi$ is the thermal diffusivity (\citealt{Ryutov_1999} p. 824); Eu is the Euler number; and $\beta$ is the plasma beta.

The laboratory plasma streams are characterized by three different scales along the x, y, and z axes, which are around 0.3, 0.1, and 1.5 cm for the laser-driven flow and 10, 3, and 100 cm for the plasma gun-driven flow. The spatial scale along the z-axis was used in Table~\ref{tabEXor}. 

As shown in Table~\ref{tabEXor}, the plasma beta values match the three considered flows perfectly and the Euler numbers are of the same order, however these values are not exactly the same. We have two laboratory flows that can each be scaled to the EXor inflow and  can reasonably correspond to  a range of parameters of the latter. Therefore, we can examine what exactly matching the EXor inflow to either the laser or plasma gun-driven flows would entail, while keeping the temperature, density, and B-field of the EXor inflow the same as in Table~\ref{tabEXor}. This leads to, for the EXor $V_{flow}$ = 290 km/s,  $Eu$ = 24, which perfectly fits the laser-driven flow case. Similarly, for the EXor $V_{flow}$ = 62 km/s, we have $Eu$ = 5.1, which corresponds to the plasma-gun driven case.

Following Ryutov's analysis, we extracted the scaling factors in space, density, and velocity between the EXor and laboratory flows. This yields the following factors, denoted with "P" and "K" for the laser-driven (PEARL) and plasma gun-driven flows (KROT), respectively. The spatial scaling parameters are as follows:
\[
a_{P} = \frac{r_{ast}}{r_{P\:lab}} = \frac{5\times 10^{6}\, {\rm [km]}}{1.5\, {\rm [cm]}} = 3.3\times 10^{11},
\]

\[
a_{K} = \frac{r_{ast}}{r_{K\:lab}} = \frac{5\times 10^{6}\, {\rm [km]}}{100\, {\rm [cm]}} = 5\times 10^{9}~.
\]

\noindent
The density scaling parameters are written as

\[
b_{P} = \frac{\rho_{ast}}{\rho_{P\:lab}} = \frac{1.1 \times 10^{-12}\, [{\rm g /cm}^{3}]}{1.4 \times 10^{-5}\, [{\rm g / cm}^{3}]} = 7.9 \times 10^{-8},
\]

\[
b_{K} = \frac{\rho_{ast}}{\rho_{K\:lab}} = \frac{1.1 \times 10^{-12}\, [{\rm g /cm}^{3}]}{1.4 \times 10^{-10}\, [{\rm g / cm}^{3}]} = 7.9 \times 10^{-3}~,
\]

\noindent
and the velocity scaling parameters are 

\[c_{P} = \frac{V_{ast}}{V_{P\:lab}} = \frac{290\, [{\rm km s}^{-1}]}{100 [{\rm km s}^{-1}]} = 2.9~,
\]

\[
c_{K} = \frac{V_{ast}}{V_{K\:lab}} = \frac{62\, [{\rm km s}^{-1}]}{23\, [{\rm km s}^{-1}]} = 2.7~.
\]

\noindent
Based on these factors, the temporal scaling $t_{ast}$ = $(a_{P} / c_{P} )t_{P\:lab}$ gives that 100 ns of the laser-driven flow is equivalent to around $1.1 \times 10^{13}$ ns ($\sim$ 3 hours) of the astrophysical flow. The magnetic field scaling $B_{ast}$ = $B_{P\:lab} c_{P} \sqrt{b_{P}}$ means that 135 kG in the laser-driven flow case corresponds to $\sim$ 110 G in the case of the EXor object. For the plasma gun-driven case $t_{ast}$ = $(a_{K} / c_{K} )t_{K\:lab}$ gives that 25 $\mu s$ is equivalent to around $9.3 \times 10^{8}$ $\mu s$ ($\sim$ 13 hours) and for 450 G in the laboratory $B_{ast}$ = $B_{K\:lab} c_{K} \sqrt{b_{K}}$ gives $\sim$ 105 G for the astrophysical case.

The similarity detailed above was made between the experiments and the EXor object detailed in \cite{Giannini_2017}. We find that a similarly good scalability can be found between our experimental measurements and the FU Orionis star, whose characterization was recently improved by \cite{labdon2020viscous}. In that case, taking the temperature value to be around 2 kK at the FU Orionis truncation radius, and assuming an averaged density of $\sim$ 10$^{13}$ cm$^{-3}$ and a velocity of $\sim$ 120 km/s for the accretion stream, we find that this stream scales that produced by the high-power laser produced plasma (detailed in Table~\ref{tabEXor}) for a $\sim$ 200 G magnetic field strength value in the astrophysical case. If the matching is done with the plasma-gun generated plasma, we find that this is possible for a magnetic field of 200 G, a density of 10$^{13}$ cm$^{-3}$ and a velocity of 30 km/s for the accretion stream. The fact that the value of the magnetic field, as inferred from the scaling, for FU Orionis is of the same order as that inferred for the EXor is consistent with the most updated knowledge (\citealt{2014prpl.conf..387A}). This fact supports the idea of identical accretion mechanisms being at play in the EXor and FUor objects.

\begin{table}
\caption{Comparison and scalability between the laser-driven (PEARL) and plasma-gun driven (KROT) plasma streams, with the EXor accretion inflow investigated in \cite{Giannini_2017}.}
\begin{center}
\begin{tabular}{cc|c|c}
 & \multicolumn{1}{c|}{\textbf{\textit{PEARL$^i$}}} & \multicolumn{1}{c|}{\textbf{\textit{KROT}}} & \multicolumn{1}{c}{\textbf{\textit{EXor}}}\tabularnewline
\hline 
\hline
Material & \multicolumn{1}{c|}{$CF_{2}$ (Teflon)} & \multicolumn{1}{c|}{$CH_{2}$} & \multicolumn{1}{c}{$H$ }\tabularnewline
Z & \multicolumn{1}{c|}{$\textbf{1}$} & \multicolumn{1}{c|}{$\textbf{1.26}$} & \multicolumn{1}{c}{$\textbf{1}$}\tabularnewline
A & \multicolumn{1}{c|}{$\textbf{17.3}$} & \multicolumn{1}{c|}{$\textbf{10.4}$} & \multicolumn{1}{c}{$\textbf{1.3}$}\tabularnewline
\hline
B $[kG]$ & \multicolumn{1}{c|}{$\textit{135}$} & \multicolumn{1}{c|}{$\textbf{0.45}$} & \multicolumn{1}{c}{\textbf{\textit{0.105}} $\bullet$}\tabularnewline
L $[cm]$ & $\textbf{1.5}$ & $\textbf{100}$ & $\mathbf{5\times 10^{11}}$\tabularnewline
n$_{e}$ $[cm^{-3}]$ & $\mathbf{5\times 10^{17}}$ & $\mathbf{10^{13}}$ & $\mathbf{5\times 10^{11}}$ $\star$\tabularnewline
%Ion density $[cm^{-3}]$ & $10^{18}$ & $6\times 10^{11}$\tabularnewline
$\rho$ $[g.cm^{-3}]$ & $1.4\times10^{-5}$ & $1.4\times10^{-10}$ & $1.1\times10^{-12}$\tabularnewline
T$_{e}$ $[kK]$ & $\textbf{18.5}$ & $\textbf{12}$ & $\textbf{12}$ $\star$\tabularnewline
T$_{i}$ $[kK]$ & $\textbf{18.5}$ & $\textbf{12}$ & $\textbf{12}$\tabularnewline
V$_{flow}$ $[km.s^{-1}]$ & $\textbf{100}$ & $\textbf{23}$ & $\textbf{180}$ $\star$\tabularnewline
C$_{S}$ $[km.s^{-1}]$ & $5.4$ & $5.9$ & $15.8$\tabularnewline
V$_{A}$ $[km.s^{-1}]$ & $100$ & $108$ & $286$\tabularnewline
l$_{e}$ $[cm]$ & $2\times10^{-5}$ & $0.14$  & $3$\tabularnewline
$\tau_{col\:e}$ $[ns]$ & $3.8\times10^{-4}$ & $3.3$  & $70$\tabularnewline
R$_{Le}$ $[cm]$ & $2.2\times10^{-5}$ & $5.3\times10^{-3}$  & $2.3\times10^{-2}$\tabularnewline
f$_{ce}$ $[s^{-1}]$ & $3.8\times10^{11}$ & $1.3\times10^{9}$  & $2.9\times10^{8}$\tabularnewline
%M$_{e}$ & $5.4$ & $320$ & $1.4 \times 10^{3}$\tabularnewline
l$_{i}$ $[cm]$ & $2.8\times10^{-5}$ & $0.12$  & $4.1$\tabularnewline
$\tau_{col\:i}$ $[ns]$ & $0.1$ & $400$  & $4.8 \times 10^{3}$\tabularnewline
R$_{Li}$ $[cm]$ & $4\times10^{-3}$  & $0.6$  & $1.1$\tabularnewline
f$_{ci}$ $[s^{-1}]$ & $1.2\times10^{7}$ & $8.3\times10^{4}$  & $1.3\times10^{5}$\tabularnewline
%M$_{i}$ & $0.04$ & $2.1$ & $42$\tabularnewline
M & $18.4$ & $3.9$ & $11.4$\tabularnewline
M$_{alf}$ & $1$ & $0.2$ & $0.6$\tabularnewline
$\tau_{\eta}$ $[ns]$ & $1.5\times10^{3}$ & $1.2\times10^{6}$  & $3.1\times10^{25}$\tabularnewline
Re$_{M}$ & $10$ & $27$ & $1.1\times10^{12}$\tabularnewline
Re & $6.3\times10^{3}$ & $650$  & $4.7\times10^{12}$\tabularnewline
Pe & $35$ & $4.7$  & $9.7\times10^{10}$\tabularnewline
Eu & $24$ & $5.1$  & $15$\tabularnewline
$\beta$ & $3.5\times10^{-3}$ & $3.6\times10^{-3}$ & $3.6\times10^{-3}$\tabularnewline
\tabularnewline
\end{tabular}
\end{center}
$^i$ The primary parameters are noted in bold, while the numbers in light are derived from these primary numbers. For the experimental streams, the primary numbers are directly measured or are deduced from simulations, as detailed in the text. For the EXor object, the B-field value is an assumed one denoted with $\bullet$ symbol, and the parameters reported by \cite{Giannini_2017} are denoted with $\star$ symbol.
\label{tabEXor}
\end{table}

\section{Discussion and conclusions}
\label{sec:conc}

Eruptive variables of the EXor/FUor type displaying powerful outbursts are still rarely observed and hence their characteristics are still not well known. Apart from their interesting characteristics, these objects could be a possible solution for the well-known the Luminosity Problem if it would confirm
that they represent a more common behaviour for YSOs than previously
thought. Indeed, the luminosity of YSOs is lower of about an order of magnitude than if the accretion proceeds at the accretion rates predicted by the standard steady-state collapse model of \cite{1977ApJ...214..488S}. This problem, originally described by \cite{1990AJ.....99..869K}, has been observationally confirmed by the Spitzer satellite survey of a number of star formation regions (\citealt{2009ApJS..181..321E}). 

We attempted to use measurements of laboratory plasma flows to complement our knowledge of these objects and in particular of the magnetic field environment in which they could take place. 
For this, we demonstrated the analogy between the accretion inflows recorded during intense outbursts of EXor/FUor objects and the laboratory plasmas created at two facilities, PEARL (high power laser) and KROT (plasma gun injector), which have different scales and plasma parameters. 

The laboratory experiments demonstrate that effective propagation of plasma is possible across B-field, which supports the proposed scenario of matter accretion not only along the magnetic field lines but also across them (see Fig. \ref{fig:accretion}). This effect can be a candidate to explain the origin of the high accretion rates of the EXor/FUor objects in comparison with standard accretion in CTTS.

The demonstrated scalability between the two laboratory plasmas and the astrophysical plasma allows us to access unknown parameters of the astrophysical system, such as the magnetic field. In particular, we have shown that we could derive the magnetic field strengths this way for the EXor object detailed in \cite{Giannini_2017} (i.e. around 100 G)  or for the FU Orionis star (\citealt{labdon2020viscous}) (i.e. around 200 G) are inconsistence with the most up-to-date understanding of such stellar objects. We note that the scaled magnetic field we infer for the astrophysical situations corresponds to average values that would be experienced by accretion streams as they propagate from the disc to the star. Lower magnetic fields are thus expected closer to the disc and correspondingly higher fields would be expected at the star surface. Since we find average values for the in-stream magnetic field of a few hundreds of G, fields in the kG range at the star surface, as found in CTTSs, would thus be reasonable. This points out that EXor/FUor-objects might not be out of the norm, but rather represent particular episodic behaviour of otherwise standard CTTSs.

Future perspectives to refine the parameters of EXor accretion streams include low-energy observations possibly tuned to more extreme cases that could then be compared with laboratory experiments tuned in the appropriate regime.
Laboratory experiments and numerical simulations are crucial for predictions on current and future high-energy observations (e.g. eROSITA, XRISM and Athena) for parameters on which no constraints can be evaluated with current data.
The topic discussed in this work is also timely  from the perspective of possible future big data surveys, which will allow us to collect multi-band observations. These include Rubin LSST in the low-energy regime along with  eROSITA, XRISM, and Athena in the high-energy regime.
Possible explanations of the lack of high statistics in the number of observed objects with accretion outbursts such as EXors could also be helped by future observational campaigns (such as the Rubin LSST ten-year survey). First, these objects could be very common, but we observed just a few of them because we need a proper cadence of observing strategy; see discussion on a possible cadence for variability in accreting stars (i.e. CTTSs and EXors) with Rubin LSST in \cite{bonito2018young}.  Second, the range of parameters that trigger the burst is too small. Numerical simulations and laboratory experiments (as those described here) are ideal to perform a wide exploration of the parameter space.
We plan to investigate these points with multi-band surveys in the near future and an active project to monitor these objects in the optical band and in X-rays with Rubin LSST and eROSITA, respectively, is ongoing\footnote{This is the topic of an approved eROSITA project, Stelzer, Giannini, and Bonito 2019}.

\begin{acknowledgements}
This work was supported by funding from the European Research Council (ERC) under the European Unions Horizon 2020 research and innovation program (Grant Agreement No. 787539). The experiments at the KROT facility were supported by the Russian Foundation for Basic Research (RFBR) in the frame of project No. 18-29-21029, and Unique Scientific Facility “Complex of Large-Scale Geophysical Facilities (KKGS)” was used. The experiments at the PEARL facility were supported by the Russian Science Foundation (RSF) in the frame of project No. 20-12-00395. The work of JIHT RAS team was done in the frame of the State Assignment (topic No. 075-00892-20-00) and partially funded by Russian Foundation for Basic Research (grant No. 18-29-21013). This work was partly done within the LABEX Plas@Par, the DIM ACAV funded by the Region Ile-de-France, and supported by Grant No. 11-IDEX- 0004-02 from ANR (France). Part of the experimental system is covered by a patent (1000183285, 2013, INPI-France). The research leading to these results is supported by Extreme Light Infrastructure Nuclear Physics (ELI-NP) Phase II, a project co-financed by the Romanian Government and European Union through the European Regional Development Fund, and by the project $\ ELI-RO-2020-23$ funded by IFA (Romania). SO acknowledges partial support from INAF-Osservatorio Astronomico di Palermo.
\end{acknowledgements}

\bibliographystyle{aa}
\bibliography{biblio}

\begin{thebibliography}{81}
\expandafter\ifx\csname natexlab\endcsname\relax\def\natexlab#1{#1}\fi

\bibitem[{Aidakina {et~al.}(2018)Aidakina, Galka, Gundorin, Gushchin, Zudin,
  Korobkov, Kostrov, Loskutov, Mogilevskiy, Priver,
  {et~al.}}]{aidakina2018simulation}
Aidakina, N., Galka, A., Gundorin, V., {et~al.} 2018, Geomagnetism and
  Aeronomy, 58, 314

\bibitem[{{Alencar} {et~al.}(2010){Alencar}, {Teixeira}, {Guimar{\~a}es},
  {McGinnis}, {Gameiro}, {Bouvier}, {Aigrain}, {Flaccomio}, \&
  {Favata}}]{2010A&A...519A..88A}
{Alencar}, S.~H.~P., {Teixeira}, P.~S., {Guimar{\~a}es}, M.~M., {et~al.} 2010,
  \aap, 519, A88

\bibitem[{{Argiroffi, C.} {et~al.}(2017){Argiroffi, C.}, {Drake, J. J.},
  {Bonito, R.}, {Orlando, S.}, {Peres, G.}, \& {Miceli, M.}}]{refId0arg2017}
{Argiroffi, C.}, {Drake, J. J.}, {Bonito, R.}, {et~al.} 2017, A\&A, 607, A14

\bibitem[{{Argiroffi, C.} {et~al.}(2007){Argiroffi, C.}, {Maggio, A.}, \&
  {Peres, G.}}]{refId02007}
{Argiroffi, C.}, {Maggio, A.}, \& {Peres, G.} 2007, A\&A, 465, L5

\bibitem[{{Argiroffi, C.} {et~al.}(2009){Argiroffi, C.}, {Maggio, A.}, {Peres,
  G.}, {Drake, J. J.}, {L\'opez-Santiago, J.}, {Sciortino, S.}, \& {Stelzer,
  B.}}]{refId02009}
{Argiroffi, C.}, {Maggio, A.}, {Peres, G.}, {et~al.} 2009, A\&A, 507, 939

\bibitem[{{Armitage}(1995)}]{1995MNRAS.274.1242A}
{Armitage}, P.~J. 1995, \mnras, 274, 1242

\bibitem[{{Armitage}(2016)}]{2016ApJ...833L..15A}
{Armitage}, P.~J. 2016, \apjl, 833, L15

\bibitem[{{Armitage}(2019)}]{2019SAAS...45....1A}
{Armitage}, P.~J. 2019, Saas-Fee Advanced Course, 45, 1

\bibitem[{{Armitage} {et~al.}(2001){Armitage}, {Livio}, \&
  {Pringle}}]{2001MNRAS.324..705A}
{Armitage}, P.~J., {Livio}, M., \& {Pringle}, J.~E. 2001, \mnras, 324, 705

\bibitem[{{Arons} \& {Lea}(1976)}]{1976ApJ...207..914A}
{Arons}, J. \& {Lea}, S.~M. 1976, \apj, 207, 914

\bibitem[{{Audard} {et~al.}(2014){Audard}, {{\'A}brah{\'a}m}, {Dunham},
  {Green}, {Grosso}, {Hamaguchi}, {Kastner}, {K{\'o}sp{\'a}l}, {Lodato},
  {Romanova}, {Skinner}, {Vorobyov}, \& {Zhu}}]{2014prpl.conf..387A}
{Audard}, M., {{\'A}brah{\'a}m}, P., {Dunham}, M.~M., {et~al.} 2014, in
  Protostars and Planets VI, ed. H.~{Beuther}, R.~S. {Klessen}, C.~P.
  {Dullemond}, \& T.~{Henning}, 387

\bibitem[{{Barbera} {et~al.}(2017){Barbera}, {Orlando}, \&
  {Peres}}]{2017A&A...600A.105B}
{Barbera}, E., {Orlando}, S., \& {Peres}, G. 2017, \aap, 600, A105

\bibitem[{{Bell} \& {Lin}(1994)}]{1994ApJ...427..987B}
{Bell}, K.~R. \& {Lin}, D.~N.~C. 1994, \apj, 427, 987

\bibitem[{bo~Tang {et~al.}(2018)bo~Tang, yue Hu, han Liang, Tao, lin Wang, Hu,
  Zhao, \& Zheng}]{Tang_2018}
bo~Tang, H., yue Hu, G., han Liang, Y., {et~al.} 2018, Plasma Physics and
  Controlled Fusion, 60, 055005

\bibitem[{Bonito {et~al.}(2018)Bonito, Hartigan, Venuti, Guarcello, Prisinzano,
  Argiroffi, Messina, Johns-Krull, Feigelson, Stauffer, Giannini, Antoniucci,
  Sciortino, Micela, Pillitteri, Fedele, Podio, Damiani, McGehee, Street,
  Gizis, Sacco, Magrini, Flaccomio, Orlando, Miceli, Stelzer, Fuchs, Chen,
  Pikuz, Frasca, Biazzo, Codella, Pastorello, Alcala', Covino, Bianchi, \&
  Nisini}]{bonito2018young}
Bonito, R., Hartigan, P., Venuti, L., {et~al.} 2018, Young Stars and their
  Variability with LSST

\bibitem[{Bonito {et~al.}(2014)Bonito, Orlando, Argiroffi, Miceli, Peres,
  Matsakos, Stehle, \& Ibgui}]{Bonito_2014}
Bonito, R., Orlando, S., Argiroffi, C., {et~al.} 2014, The Astrophysical
  Journal, 795, L34

\bibitem[{{Bonito} {et~al.}(2010{\natexlab{a}}){Bonito}, {Orlando}, {Miceli},
  {Eisl{\"o}ffel}, {Peres}, \& {Favata}}]{bom10}
{Bonito}, R., {Orlando}, S., {Miceli}, M., {et~al.} 2010{\natexlab{a}}, \aap,
  517, A68

\bibitem[{{Bonito} {et~al.}(2010{\natexlab{b}}){Bonito}, {Orlando}, {Peres},
  {Eisl{\"o}ffel}, {Miceli}, \& {Favata}}]{bop10}
{Bonito}, R., {Orlando}, S., {Peres}, G., {et~al.} 2010{\natexlab{b}}, \aap,
  511, A42

\bibitem[{{Bonnell} \& {Bastien}(1992)}]{1992ApJ...401L..31B}
{Bonnell}, I. \& {Bastien}, P. 1992, \apjl, 401, L31

\bibitem[{{Bouvier} {et~al.}(2007){Bouvier}, {Alencar}, {Boutelier},
  {Dougados}, {Balog}, {Grankin}, {Hodgkin}, {Ibrahimov}, {Kun}, {Magakian}, \&
  {Pinte}}]{2007A&A...463.1017B}
{Bouvier}, J., {Alencar}, S.~H.~P., {Boutelier}, T., {et~al.} 2007, \aap, 463,
  1017

\bibitem[{Braginskii(1965)}]{Braginskii_1965}
Braginskii, S.~I. 1965, Reviews of Plasma Physics, 1, 205

\bibitem[{Bruneteau {et~al.}(1970)Bruneteau, Fabre, Lamain, \&
  Vasseur}]{doi:10.1063/1.1693157}
Bruneteau, J., Fabre, E., Lamain, H., \& Vasseur, P. 1970, The Physics of
  Fluids, 13, 1795

\bibitem[{{Burdonov, K.} {et~al.}(2020){Burdonov, K.}, {Revet, G.}, {Bonito,
  R.}, {Argiroffi, C.}, {B\'eard, J.}, {Bolan\~os, S.}, {Cerchez, M.}, {Chen,
  S. N.}, {Ciardi, A.}, {Espinosa, G.}, {Filippov, E.}, {Pikuz, S.},
  {Rodriguez, R.}, {Sm\'{\i}d, M.}, {Starodubtsev, M.}, {Willi, O.}, {Orlando,
  S.}, \& {Fuchs, J.}}]{burdonov2020laboratory}
{Burdonov, K.}, {Revet, G.}, {Bonito, R.}, {et~al.} 2020, A\&A, 642, A38

\bibitem[{{Camenzind}(1990)}]{1990RvMA....3..234C}
{Camenzind}, M. 1990, Reviews in Modern Astronomy, 3, 234

\bibitem[{Ciardi {et~al.}(2007)Ciardi, Lebedev, Frank, Blackman, Chittenden,
  Jennings, Ampleford, Bland, Bott, Rapley, {et~al.}}]{ciardi2007evolution}
Ciardi, A., Lebedev, S., Frank, A., {et~al.} 2007, Physics of Plasmas, 14,
  056501

\bibitem[{Ciardi {et~al.}(2013)Ciardi, Vinci, Fuchs, Albertazzi, Riconda,
  P{\'e}pin, \& Portugall}]{ciardi2013astrophysics}
Ciardi, A., Vinci, T., Fuchs, J., {et~al.} 2013, Physical review letters, 110,
  025002

\bibitem[{{Coffaro} {et~al.}(2020){Coffaro}, {Stelzer}, {Orlando}, {Hall},
  {Metcalfe}, {Wolter}, {Mittag}, {Sanz-Forcada}, {Schneider}, \&
  {Ducci}}]{2020A&A...636A..49C}
{Coffaro}, M., {Stelzer}, B., {Orlando}, S., {et~al.} 2020, \aap, 636, A49

\bibitem[{Colombier {et~al.}(2005)Colombier, Combis, Bonneau, Le~Harzic, \&
  Audouard}]{PhysRevB.71.165406}
Colombier, J.~P., Combis, P., Bonneau, F., Le~Harzic, R., \& Audouard, E. 2005,
  Phys. Rev. B, 71, 165406

\bibitem[{{Colombo} {et~al.}(2019){Colombo}, {Orlando}, {Peres}, {Reale},
  {Argiroffi}, {Bonito}, {Ibgui}, \& {Stehl{\'e}}}]{2019A&A...624A..50C}
{Colombo}, S., {Orlando}, S., {Peres}, G., {et~al.} 2019, \aap, 624, A50

\bibitem[{{Cranmer}(2009)}]{2009ApJ...706..824C}
{Cranmer}, S.~R. 2009, \apj, 706, 824

\bibitem[{{D'Angelo} \& {Spruit}(2012)}]{2012MNRAS.420..416D}
{D'Angelo}, C.~R. \& {Spruit}, H.~C. 2012, \mnras, 420, 416

\bibitem[{Donati {et~al.}(2011{\natexlab{a}})Donati, Bouvier, Walter, Gregory,
  Skelly, Hussain, Flaccomio, Argiroffi, Grankin, Jardine, Ménard, Dougados,
  Romanova, \& the MaPP~collaboration}]{10.1111/j.1365-2966.2010.18069.x}
Donati, J.-F., Bouvier, J., Walter, F.~M., {et~al.} 2011{\natexlab{a}}, Monthly
  Notices of the Royal Astronomical Society, 412, 2454

\bibitem[{Donati {et~al.}(2011{\natexlab{b}})Donati, Gregory, Montmerle,
  Maggio, Argiroffi, Sacco, Hussain, Kastner, Alencar, Audard, Bouvier,
  Damiani, Güdel, Huenemoerder, \& Wade}]{10.1111/j.1365-2966.2011.19366.x}
Donati, J.-F., Gregory, S.~G., Montmerle, T., {et~al.} 2011{\natexlab{b}},
  Monthly Notices of the Royal Astronomical Society, 417, 1747

\bibitem[{Donati {et~al.}(2005)Donati, Paletou, Bouvier, \&
  Ferreira}]{donati2005direct}
Donati, J.-F., Paletou, F., Bouvier, J., \& Ferreira, J. 2005, nature, 438, 466

\bibitem[{{Evans} {et~al.}(2009){Evans}, {Dunham}, {J{\o}rgensen}, {Enoch},
  {Mer{\'\i}n}, {van Dishoeck}, {Alcal{\'a}}, {Myers}, {Stapelfeldt}, {Huard},
  {Allen}, {Harvey}, {van Kempen}, {Blake}, {Koerner}, {Mundy}, {Padgett}, \&
  {Sargent}}]{2009ApJS..181..321E}
{Evans}, Neal~J., I., {Dunham}, M.~M., {J{\o}rgensen}, J.~K., {et~al.} 2009,
  \apjs, 181, 321

\bibitem[{Giannini {et~al.}(2017)Giannini, Antoniucci, Lorenzetti, Munari,
  Causi, Manara, Nisini, Arkharov, Dallaporta, Paola, Giunta, Harutyunyan,
  Klimanov, Marchetti, Righetti, Rossi, Strafella, \& Testa}]{Giannini_2017}
Giannini, T., Antoniucci, S., Lorenzetti, D., {et~al.} 2017, The Astrophysical
  Journal, 839, 112

\bibitem[{Green {et~al.}(2013)Green, Robertson, Baek, Pooley, Pak, Im, Lee,
  Jeon, Choi, \& Meschiari}]{green2013variability}
Green, J.~D., Robertson, P., Baek, G., {et~al.} 2013, The Astrophysical
  Journal, 764, 22

\bibitem[{{Gullbring} {et~al.}(1996){Gullbring}, {Barwig}, {Chen}, {Gahm}, \&
  {Bao}}]{1996A&A...307..791G}
{Gullbring}, E., {Barwig}, H., {Chen}, P.~S., {Gahm}, G.~F., \& {Bao}, M.~X.
  1996, \aap, 307, 791

\bibitem[{Gushchin {et~al.}(2018)Gushchin, Korobkov, Terekhin, Strikovskiy,
  Gundorin, Zudin, Aidakina, \& Nikolenko}]{gushchin2018laboratory}
Gushchin, M.~E., Korobkov, S.~V., Terekhin, V.~A., {et~al.} 2018, JETP Letters,
  108, 391

\bibitem[{{Hartmann}(2008)}]{2008apsf.book.....H}
{Hartmann}, L. 2008, {Accretion Processes in Star Formation} (Cambridge
  University Press)

\bibitem[{Higginson {et~al.}(2017)Higginson, Revet, Khiar, Béard, Blecher,
  Borghesi, Burdonov, Chen, Filippov, Khaghani, Naughton, Pépin, Pikuz,
  Portugall, Riconda, Riquier, Ryazantsev, Skobelev, Soloviev, Starodubtsev,
  Vinci, Willi, Ciardi, \& Fuchs}]{HIGGINSON201748}
Higginson, D., Revet, G., Khiar, B., {et~al.} 2017, High Energy Density
  Physics, 23, 48

\bibitem[{Hillas(1984)}]{doi:10.1146/annurev.aa.22.090184.002233}
Hillas, A.~M. 1984, Annual Review of Astronomy and Astrophysics, 22, 425

\bibitem[{Ivanov {et~al.}(2006)}]{ivanov2006}
Ivanov, V.~V. {et~al.} 2006, Physics of Plasmas, 13, 012704

\bibitem[{Johns-Krull(2007)}]{Johns_Krull_2007}
Johns-Krull, C.~M. 2007, The Astrophysical Journal, 664, 975

\bibitem[{Johns-Krull {et~al.}(2009)Johns-Krull, Greene, Doppmann, \&
  Covey}]{Johns_Krull_2009}
Johns-Krull, C.~M., Greene, T.~P., Doppmann, G.~W., \& Covey, K.~R. 2009, The
  Astrophysical Journal, 700, 1440

\bibitem[{Kastner {et~al.}(2002)Kastner, Huenemoerder, Schulz, Canizares, \&
  Weintraub}]{Kastner_2002}
Kastner, J.~H., Huenemoerder, D.~P., Schulz, N.~S., Canizares, C.~R., \&
  Weintraub, D.~A. 2002, The Astrophysical Journal, 567, 434

\bibitem[{{Kenyon} {et~al.}(1990){Kenyon}, {Hartmann}, {Strom}, \&
  {Strom}}]{1990AJ.....99..869K}
{Kenyon}, S.~J., {Hartmann}, L.~W., {Strom}, K.~M., \& {Strom}, S.~E. 1990,
  \aj, 99, 869

\bibitem[{Khiar {et~al.}(2019)Khiar, Revet, Ciardi, Burdonov, Filippov,
  B{\'e}ard, Cerchez, Chen, Gangolf, Makarov, {et~al.}}]{khiar2019laser}
Khiar, B., Revet, G., Ciardi, A., {et~al.} 2019, Physical Review Letters, 123,
  205001

\bibitem[{{Koenigl}(1991)}]{1991ApJ...370L..39K}
{Koenigl}, A. 1991, \apjl, 370, L39

\bibitem[{{Koldoba} {et~al.}(2002){Koldoba}, {Romanova}, {Ustyugova}, \&
  {Lovelace}}]{2002ApJ...576L..53K}
{Koldoba}, A.~V., {Romanova}, M.~M., {Ustyugova}, G.~V., \& {Lovelace},
  R.~V.~E. 2002, \apjl, 576, L53

\bibitem[{{Korobkov} {et~al.}(2019){Korobkov}, {Gushchin}, {Gundorin}, {Zudin},
  {Aidakina}, {Strikovskiy}, \& {Nikolenko}}]{2019TePhL..45..228K}
{Korobkov}, S.~V., {Gushchin}, M.~E., {Gundorin}, V.~I., {et~al.} 2019,
  Technical Physics Letters, 45, 228

\bibitem[{{Kulkarni} \& {Romanova}(2008)}]{2008MNRAS.386..673K}
{Kulkarni}, A.~K. \& {Romanova}, M.~M. 2008, \mnras, 386, 673

\bibitem[{Labdon {et~al.}(2020)Labdon, Kraus, Davies, Kreplin, Monnier,
  Bouquin, Anugu, Brummelaar, Setterholm, Gardener, Ennis, Lanthermann,
  Schaefer, \& Laws}]{labdon2020viscous}
Labdon, A., Kraus, S., Davies, C.~L., {et~al.} 2020, Viscous Heating and
  Boundary Layer Accretion in the Disk of Outbursting Star FU Orionis

\bibitem[{{Lasota}(2001)}]{2001NewAR..45..449L}
{Lasota}, J.-P. 2001, \nar, 45, 449

\bibitem[{{Lodato} \& {Clarke}(2004)}]{2004MNRAS.353..841L}
{Lodato}, G. \& {Clarke}, C.~J. 2004, \mnras, 353, 841

\bibitem[{Lozhkarev {et~al.}(2007)Lozhkarev, Freidman, Ginzburg, Katin,
  Khazanov, Kirsanov, Luchinin, Mal{\'}shakov, Martyanov, Palashov, Poteomkin,
  Sergeev, Shaykin, \& Yakovlev}]{Lozhkarev_2007}
Lozhkarev, V.~V., Freidman, G.~I., Ginzburg, V.~N., {et~al.} 2007, Laser
  Physics Letters, 4, 421

\bibitem[{Marshall(1960)}]{Marshall_1960}
Marshall, J. 1960, Phys. Fluids, 3, 134

\bibitem[{{Mu\~{n}oz Caro} \& Escribano(2018)}]{munoz}
{Mu\~{n}oz Caro}, G.~M. \& Escribano, R. 2018, {Laboratory Astrophysics} (New
  York: Springer, 2018)

\bibitem[{{Nayakshin} \& {Lodato}(2012)}]{2012MNRAS.426...70N}
{Nayakshin}, S. \& {Lodato}, G. 2012, \mnras, 426, 70

\bibitem[{{Orlando} {et~al.}(2011){Orlando}, {Reale}, {Peres}, \&
  {Mignone}}]{2011MNRAS.415.3380O}
{Orlando}, S., {Reale}, F., {Peres}, G., \& {Mignone}, A. 2011, \mnras, 415,
  3380

\bibitem[{Perevalov {et~al.}(2020)Perevalov, Burdonov, Kotov, Romanovskiy,
  Soloviev, Starodubtsev, Golovanov, Ginzburg, Kochetkov, Korobeinikova,
  Kuz'min, Shaikin, Shaykin, Yakovlev, Khazanov, \& Kostyukov}]{Perevalov_2020}
Perevalov, S.~E., Burdonov, K.~F., Kotov, A.~V., {et~al.} 2020, Plasma Physics
  and Controlled Fusion, 62, 094004

\bibitem[{Plechaty {et~al.}(2013)Plechaty, Presura, \&
  Esaulov}]{PhysRevLett.111.185002}
Plechaty, C., Presura, R., \& Esaulov, A.~A. 2013, Phys. Rev. Lett., 111,
  185002

\bibitem[{Remington {et~al.}(2006)Remington, Drake, \&
  Ryutov}]{RevModPhys.78.755}
Remington, B.~A., Drake, R.~P., \& Ryutov, D.~D. 2006, Rev. Mod. Phys., 78, 755

\bibitem[{Revet(2018)}]{revet:tel-02100492}
Revet, G. 2018, Theses, {Universit{\'e} Paris Saclay (COmUE)}

\bibitem[{{Revet} {et~al.}(2017){Revet}, {Chen}, {Bonito}, {Khiar}, {Filippov},
  {Argiroffi}, {Higginson}, {Orlando}, {B{\'e}ard}, {Blecher}, {Borghesi},
  {Burdonov}, {Khaghani}, {Naughton}, {P{\'e}pin}, {Portugall}, {Riquier},
  {Rodriguez}, {Ryazantsev}, {Skobelev}, {Soloviev}, {Willi}, {Pikuz},
  {Ciardi}, \& {Fuchs}}]{2017SciA....3E0982R}
{Revet}, G., {Chen}, S.~N., {Bonito}, R., {et~al.} 2017, Science Advances, 3

\bibitem[{Richardson(2019)}]{NRL_pf}
Richardson, A. 2019, NRL Plasma Formulary

\bibitem[{{Romanova} {et~al.}(2002){Romanova}, {Ustyugova}, {Koldoba}, \&
  {Lovelace}}]{2002ApJ...578..420R}
{Romanova}, M.~M., {Ustyugova}, G.~V., {Koldoba}, A.~V., \& {Lovelace},
  R.~V.~E. 2002, \apj, 578, 420

\bibitem[{{Romanova} {et~al.}(2004){Romanova}, {Ustyugova}, {Koldoba}, \&
  {Lovelace}}]{2004ApJ...610..920R}
{Romanova}, M.~M., {Ustyugova}, G.~V., {Koldoba}, A.~V., \& {Lovelace},
  R.~V.~E. 2004, \apj, 610, 920

\bibitem[{{Romanova} {et~al.}(2003){Romanova}, {Ustyugova}, {Koldoba}, {Wick},
  \& {Lovelace}}]{2003ApJ...595.1009R}
{Romanova}, M.~M., {Ustyugova}, G.~V., {Koldoba}, A.~V., {Wick}, J.~V., \&
  {Lovelace}, R.~V.~E. 2003, \apj, 595, 1009

\bibitem[{Ryutov {et~al.}(1999)Ryutov, Drake, Kane, Liang, Remington, \&
  Wood-Vasey}]{Ryutov_1999}
Ryutov, D., Drake, R.~P., Kane, J., {et~al.} 1999, The Astrophysical Journal,
  518, 821

\bibitem[{Ryutov(2018)}]{doi:10.1063/1.5042254}
Ryutov, D.~D. 2018, Physics of Plasmas, 25, 100501

\bibitem[{Ryutov {et~al.}(2000)Ryutov, Drake, \& Remington}]{Ryutov_2000}
Ryutov, D.~D., Drake, R.~P., \& Remington, B.~A. 2000, The Astrophysical
  Journal Supplement Series, 127, 465

\bibitem[{{Safier}(1998)}]{1998ApJ...494..336S}
{Safier}, P.~N. 1998, \apj, 494, 336

\bibitem[{{Sanz-Forcada} {et~al.}(2013){Sanz-Forcada}, {Stelzer}, \&
  {Metcalfe}}]{2013A&A...553L...6S}
{Sanz-Forcada}, J., {Stelzer}, B., \& {Metcalfe}, T.~S. 2013, \aap, 553, L6

\bibitem[{{Scharlemann}(1978)}]{1978ApJ...219..617S}
{Scharlemann}, E.~T. 1978, \apj, 219, 617

\bibitem[{{Shu}(1977)}]{1977ApJ...214..488S}
{Shu}, F.~H. 1977, \apj, 214, 488

\bibitem[{{Shu} {et~al.}(1994){Shu}, {Najita}, {Ruden}, \&
  {Lizano}}]{1994ApJ...429..797S}
{Shu}, F.~H., {Najita}, J., {Ruden}, S.~P., \& {Lizano}, S. 1994, \apj, 429,
  797

\bibitem[{Soloviev {et~al.}(2017)Soloviev, Burdonov, Chen, Eremeev,
  Korzhimanov, Pokrovskiy, Pikuz, Revet, Sladkov, Ginzburg, Khazanov, Kuzmin,
  Osmanov, Shaikin, Shaykin, Yakovlev, Pikuz, Starodubtsev, \&
  Fuchs}]{Soloviev2017}
Soloviev, A., Burdonov, K., Chen, S.~N., {et~al.} 2017, Scientific Reports, 7,
  12144

\bibitem[{Stauffer {et~al.}(2014)Stauffer, Cody, Baglin, Alencar, Rebull,
  Hillenbrand, Venuti, Turner, Carpenter, Plavchan, Findeisen, Carey, Terebey,
  Morales-Calder{\'{o}}n, Bouvier, Micela, Flaccomio, Song, Gutermuth,
  Hartmann, Calvet, Whitney, Barrado, Vrba, Covey, Herbst, Furesz, Aigrain, \&
  Favata}]{Stauffer_2014}
Stauffer, J., Cody, A.~M., Baglin, A., {et~al.} 2014, The Astronomical Journal,
  147, 83

\bibitem[{Sucov {et~al.}(1967)Sucov, Pack, Phelps, \&
  Engelhardt}]{doi:10.1063/1.1762404}
Sucov, E.~W., Pack, J.~L., Phelps, A.~V., \& Engelhardt, A.~G. 1967, The
  Physics of Fluids, 10, 2035

\bibitem[{{Vorobyov} \& {Basu}(2005)}]{2005ApJ...633L.137V}
{Vorobyov}, E.~I. \& {Basu}, S. 2005, \apjl, 633, L137

\end{thebibliography}

\end{document}